\documentclass[a4paper,11pt]{article}
\pdfoutput=1 
\usepackage{jcappub} 

\usepackage{aasmacros} 

\usepackage{caption}
\usepackage{courier}
\usepackage{natbib}
\usepackage[utf8]{inputenc} 
\usepackage{hyperref}       
\usepackage{url}            
\usepackage[dvipsnames]{xcolor}
\usepackage{subfigure}
\usepackage{caption}
\usepackage{subcaption}
\usepackage{graphicx}
\usepackage{lineno}
\usepackage{multirow}
\usepackage{amsmath}
\usepackage{amssymb}
\usepackage{lipsum}  
\usepackage{verbatim}
\usepackage[normalem]{ulem}

\newcommand{\Msun}{{\rm M}_\odot}

\newcommand{\de}{{\rm d}}

\title{Dark matter decay signals in cosmic filaments}

\author[a,b,c,d]{Elena Pinetti,}
\author[e]{Evan Vienneau,}
\author[e,f]{and Nassim Bozorgnia}

\affiliation[a]{Center for Computational Astrophysics, Flatiron Institute, New York, NY 10010, USA}
\affiliation[b]{Fermi National Accelerator Laboratory, Theoretical Astrophysics Department, Batavia, IL, 60510, USA}
\affiliation[c]{University of Chicago, Kavli Institute for Cosmological Physics, Chicago, IL 60637, USA}
\affiliation[d]{Emmy Noether Fellow, Perimeter Institute for Theoretical Physics, 31 Caroline Street N., Waterloo, Ontario, N2L 2Y5, Canada}
\affiliation[e]{Department of Physics, University of Alberta, CCIS 4-181, Edmonton, Alberta T6G 2E1, Canada}
\affiliation[f]{Theoretical Physics Institute, University of Alberta, CCIS 4-181, Edmonton, Alberta T6G 2E1, Canada}

\emailAdd{epinetti@flatironinstitute.org}
\emailAdd{eviennea@ualberta.ca}
\emailAdd{nbozorgnia@ualberta.ca}

\abstract{Cosmic filaments form the backbone of the cosmic web, yet their properties and evolution remain uncertain. Using the EAGLE hydrodynamical simulations, we investigate the dark matter density profiles in filaments and their implications for dark matter decay signals. We show that GeV-scale dark matter particles decaying into electron-positron pairs can produce detectable radio synchrotron emission. 
By leveraging stacked filament radio data, we place stringent constraints on the dark matter decay lifetime, improving existing limits by up to two orders of magnitude for strong filamentary magnetic fields.}

\begin{document}
\maketitle
\flushbottom

\section{Introduction}
\label{sec:introduction}

Filaments in the Universe are fundamental components of the cosmic web, the large-scale structure that permeates the cosmos. These vast, thread-like formations, composed mainly of dark matter (DM), gas, and galaxies, serve as the connective bridges between galaxy clusters, spanning few to tens of Mpc~\cite{Bond:1995yt, White:1987yr}. They play a crucial role in the evolution and distribution of matter throughout the Universe, channeling gas into galaxies and driving large-scale dynamics. Therefore, understanding the properties and behavior of these filaments offers insights into the processes governing cosmic structure formation and the underlying physics that shape the Universe. 

Theories regarding the non-uniform, web-like distribution of matter in the Universe arose from studies of gravitational collapse and structure formation~\cite{Peebles:1980yev, Zeldovich:1982zz, White:1987yr, Bond:1995yt}. Subsequent analysis of galaxy surveys provided the first evidence for the existence of the cosmic web by analyzing the positions of galaxies~\cite{Bahcall:1988ch, Gott:1988rj}. Later confirmation came from large galaxy redshift surveys such as the Sloan Digital Sky Survey~\cite{Margon:1998vu} and the 2dF Galaxy Redshift Survey~\cite{2dFGRS:2005yhx}. Observations of the visible matter content of the cosmic web have allowed for the study of galaxy characteristics~\cite{Dubois:2014lxa, Chen:2015oqa, Kuutma:2017yvb, Laigle:2017byb, Sarron:2019cyf, Bonjean:2019hnz} and the gas content~\cite{Bonjean:2017okn, Tanimura:2019uxm} in filaments.

The first robust detection of stacked radio emission from large filaments ($>3$~Mpc) was reported in ref.~\cite{Vernstrom:2021hru}. The study shows that the observed radio emission is 40 times stronger than predicted by simulations of shocked intergalactic gas. Their analysis rules out unsubtracted point sources and instrumental or systematic effects as possible explanations.
An intriguing possibility is that this discrepancy arises from a DM contribution. DM accounts for $\sim 84\%$ of the matter content in the Universe~\cite{Planck:2018vyg}
and is expected to make up the majority of the mass in filaments.  A recent study \cite{Dunsky:2025pvd} demonstrated that cosmological filaments provide a unique environment to probe the intriguing scenario in which DM particles decay into gravitons, which subsequently convert into gamma rays in the presence of nanoGauss filamentary magnetic fields. Moreover, ref.~\cite{Vernstrom:2021hru} shows that DM candidates with a mass of 5–10 GeV could produce a radio signal comparable to observations if they decay into electron-positron pairs, which subsequently generate radio waves through synchrotron radiation. The main limitation of the analysis done in ref.~\cite{Vernstrom:2021hru} is the assumption of a constant DM density profile in filaments.   

A realistic description of the expected DM decay signals from cosmic filaments requires first and foremost, characterizing the distribution of DM within filaments. Large-scale hydrodynamical simulations including EAGLE~\cite{EAGLE:2017, Schaye:2014tpa, Crain:2015poa}, BAHAMAS~\cite{McCarthy:2016mry}, Illustris-TNG~\cite{2018MNRAS.473.4077P, 2018MNRAS.475..676S}, MillenniumTNG~\cite{2023MNRAS.524.2539P}, and FLAMINGO~\cite{Schaye:2023jqv, Kugel:2023wte}, provide a valuable framework for investigating these filamentary structures. Their extensive simulation volumes—ranging from 100 Mpc in EAGLE and BAHAMAS to 205 Mpc in TNG300, 500 Mpc in MillenniumTNG, and up to 1–2.8 Gpc in FLAMINGO—allow for the detailed study of the DM distribution within inter-cluster filaments. The density profiles of DM, gas and stars in filaments have been extracted from the  EAGLE~\cite{Bahe:2025kua} and
Illustris-TNG~\cite{Galarraga-Espinosa:2021kbe} simulations. However, the majority of work studying filaments within simulations have focused on characterizing the properties of galaxies/halos within and around filaments~\cite{Hahn:2007ui, Codis:2012ep, Trowland:2012pi, Laigle:2013tsa, Borzyszkowski:2016kmi, Gheller:2016knp, GaneshaiahVeena:2019arz,Banfi:2021nyj, Hodgson_2021, May:2022gus, Lokken:2022omq, Jhee:2022lup, Das:2023aje, Hasan:2023ujm, Hunde:2024wic} and the distribution of gas within filaments~\cite{Martizzi:2018iik, Gouin:2022kcr, Li:2022zwu, Vurm:2023yze}. Additionally, the DM distribution in filaments has been 
characterized using theoretical frameworks for the dynamics of the formation of the cosmic web~\cite{Feldbrugge:2017ivf, Feldbrugge:2024wsb, Feldbrugge:2022npw}.

In this work, we use the EAGLE (``Evolution and Assembly of GaLaxies and their Environments") hydrodynamical simulations~\cite{EAGLE:2017, Schaye:2014tpa, Crain:2015poa} to study the DM densities in cosmic filaments and their implications for indirect detection signals from DM decay. The EAGLE project is a suite of cosmological, hydrodynamical simulations calibrated to match observed distributions of stellar masses and sizes of low-redshift galaxies. We first identify the cosmic filaments in the simulation volume, and then extract the DM density profile along the radius and length of the filaments. Using the average DM density profiles extracted from the simulations, we compute the DM decay signals into electron-positron pairs and discuss their implications for the DM lifetime. We also explore whether DM decay could account for the recently observed radio excess \cite{Vernstrom:2021hru} through the production of electron-positron pairs emitting synchrotron radiation in filamentary magnetic fields.

The paper is structured as follows. Section \ref{sec:simulations} describes the simulations used in this work, with a particular focus on filament selection. Section~\ref{sec:density} outlines the DM density profile within filaments. In section~\ref{sec:DM} we discuss the signals expected from DM decay in filaments and in section~\ref{sec:results} we present our main results. Finally, section~\ref{sec:conclusions} discusses our key conclusions and future prospects.

\section{Simulations and filament selection}
\label{sec:simulations}

In this work we use the public data release of the cosmological, hydrodynamical simulations of the EAGLE  project~\cite{Schaye:2014tpa, Crain:2015poa, EAGLE:2017}. The EAGLE project utilizes cutting-edge numerical methods and subgrid models to simulate a wide range of astrophysical processes essential for galaxy formation in a standard $\Lambda$CDM ($\Lambda$ Cold Dark Matter) universe. The simulations were performed using a modified version of the \texttt{GADGET-3} tree-Smoothed Particle Hydrodynamics (SPH) code~\cite{Springel:2005mi}. The galaxy formation subgrid model includes star formation, stellar evolution, radiative cooling,  supernovae and AGN feedback, as well as the seeding, growth and feedback from black holes~\cite{Crain:2015poa}. The numerical parameters for the subgrid model were calibrated to reproduce the $z = 0$ galaxy stellar mass
function, galaxy sizes, and the relation between the black hole mass and stellar mass.

We use the \texttt{Reference} simulation volume, which has a side length of 100~comoving Mpc (cMpc) at intermediate resolution, with an initial baryonic particle mass of $m_g = 1.8 \times 10^6~\Msun$, DM particle mass of $m_{\rm DM} = 9.7 \times 10^6~\Msun$, and a Plummer-equivalent gravitational softening length of $\epsilon_p = 0.7$~kpc~\cite{Power:2002sw}. This resolution level (both in DM particle mass and softening length) is sufficient for the purposes of our study, considering the large scales and masses of cosmic filaments.  The simulations adopt the Planck 2013 cosmological parameters~\cite{Planck:2013pxb}: $\Omega_m=0.307$, $\Omega_\Lambda=0.693$, $\Omega_b=0.04825$, $h=0.6777$, $\sigma_8=0.8288$, and $n_s=0.9611$.

In order to select the filaments in the simulations, we first need to identify a set of galaxies that trace those filaments. We select these tracer galaxies by requiring that their stellar mass is $\log_{10} (M_{*}/\Msun) \geq 9$, taking into account observational limits~\cite{Brinchmann_2004, Taylor_2011}. With this criterion, we obtain a sample of 13,200 tracer galaxies. To extract the filamentary structure from the cosmic web, we use the Discrete Persistent Structures Extractor (DisPerSE) software~\cite{Sousbie:2010fp, Sousbie:2010fn} applied to our set of tracer galaxies at $z=0$. 
The DisPerSe formalism is built upon Discrete Morse theory, which allows one to examine the topology of a smooth manifold by analyzing the critical points and integral lines of a function defined over the manifold. For our purposes, this framework can be used to extract filaments from the topology of a density field defined by a set of tracers. The detailed user's guide regarding the use of DisPerSE to extract filaments from the cosmic web is provided in ref.~\cite{Galarraga-Espinosa:2023zmv}.

To begin, we compute the local tracer number density at each galaxy position using the Delaunay tessellation field estimator~\cite{Schaap_2000, vandeWeygaert:2007ze}, which is implemented in DisPerSE. The resulting density field is then smoothed by averaging the density at each vertex of the Delaunay tessellation (i.e.~position of each tracer) with that of its direct neighbours a single time. This minimizes the effect of shot noise by removing the small-scale, insignificant features.   
Next, we determine the location of critical points in the density field and then extract the filamentary structures. Following refs.~\cite{Galarraga-Espinosa:2020rhp, Galarraga-Espinosa:2021kbe}, filaments are defined as segments connecting maximum critical points to saddle critical points, effectively tracing the ridges of the density field. We thus obtain a catalogue of filaments, each defined by a collection of coordinates tracing out the spine.  
A final smoothing is applied by averaging the coordinates which define each filament spine with its nearest neighbours~\cite{Galarraga-Espinosa:2023zmv}. 
Note that filaments may be defined differently depending on the specific filament extraction method that is being used~\cite{Libeskind:2017tun}. However, since we are interested in the average density profile of all filaments, our results are not sensitive to the exact filament definition.

A key parameter in DisPerSE associated with each filament is \emph{persistence}, which quantifies the ratio of densities between the two critical points defining the filament’s endpoints. This parameter reflects the topological significance of the filament within the cosmic web. To ensure a meaningful selection, we include only filaments with persistence values above a specified threshold, defined in terms of the number of standard deviations, $\sigma$, above the mean persistence value. Choosing an appropriate threshold is essential for an accurate reconstruction of the cosmic web. To this end, we explore a range of thresholds in order to determine the optimal value.

In order to calibrate the $\sigma$ parameter, we utilize the concepts of \emph{completeness} and \emph{purity}~\cite{Cornwell:2023ntz,hasan2024}. Following ref.~\cite{Cornwell:2023ntz}, we define completeness as the fraction of clusters that contain a maximum-density critical point (CP$_{\rm max}$) within their virial\footnote{Virial radius and mass are defined as those corresponding to a sphere containing a mean matter density 200 times the critical density of the Universe.} radius, $R_{\rm 200}$. Purity is defined as the fraction of CP$_{\rm max}$'s that lie within the $R_{200}$ of any cluster. Here, we define clusters as halos with virial mass, $\log_{10}(M_{\rm 200}/\Msun) \geq 13$.  
We aim to simultaneously maximize these two quantities such that the resulting filamentary structure is \textit{complete} in that it accurately reconstructs the majority of the cosmic web, whilst remaining \textit{pure} in that there is minimal contamination by spurious features. A higher persistence threshold results in an increase in the purity and a simultaneous decrease in completeness. 

\begin{figure}[t]
    \centering
    \includegraphics[width=0.7\linewidth]{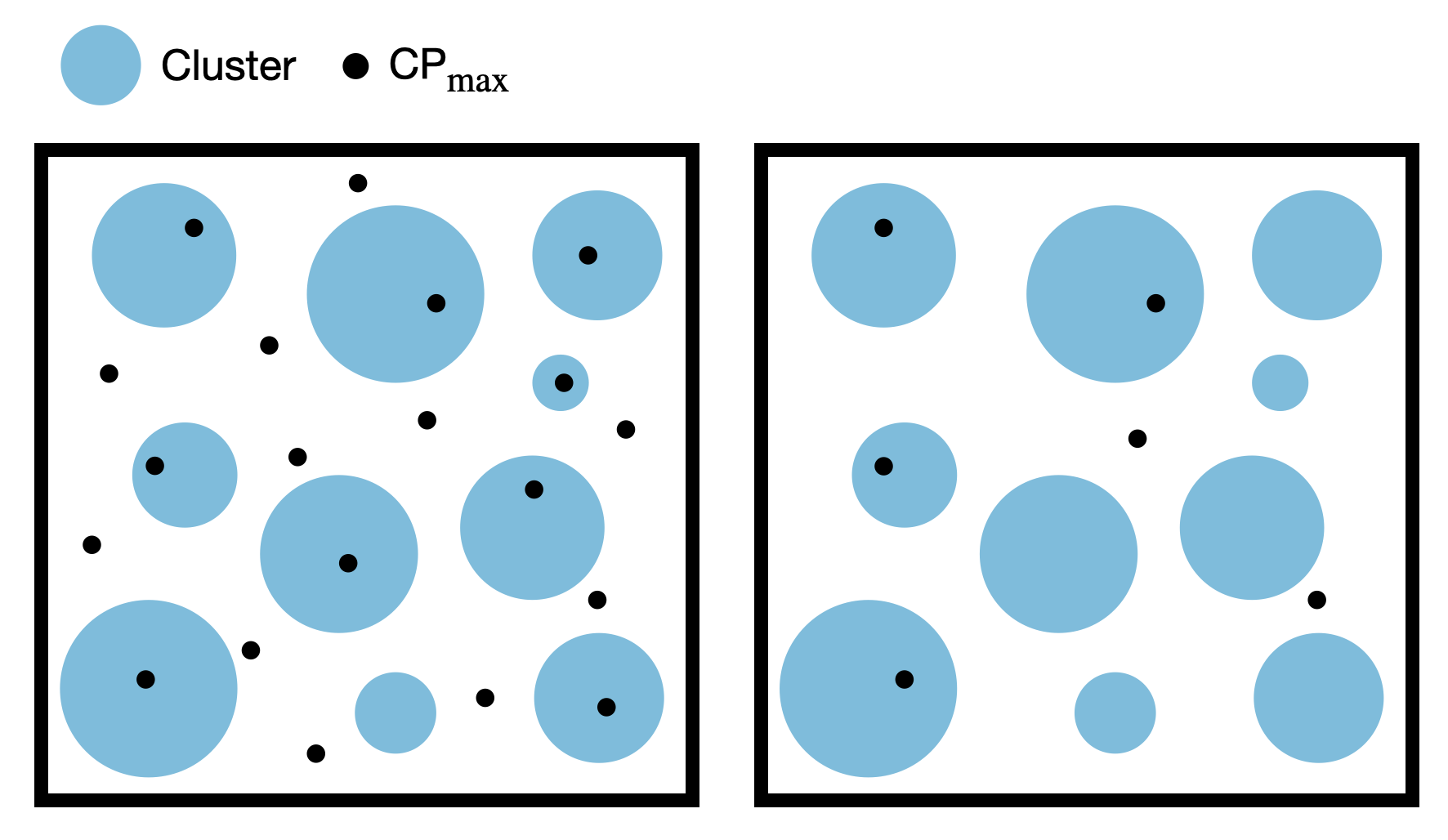}\hspace{5pt}
    \caption{A schematic diagram illustrating the concepts of purity and completeness. The blue circles and black dots show the clusters and maximum-density critical points (CP$_{\text{max}}$) distributed in a box representing the simulation volume. The left  box shows the distribution of CP$_{\text{max}}$'s for a small $\sigma$ persistence threshold, which results in more CP$_{\text{max}}$'s, a higher completeness and a lower purity. The right  box shows the distribution of CP$_{\text{max}}$'s for a large $\sigma$ persistence threshold, resulting in less CP$_{\text{max}}$'s, a lower completeness and a higher purity.}
    \label{purity_completeness}
\end{figure}

Figure~\ref{purity_completeness} demonstrates the relationship between purity, completeness and the $\sigma$ parameter of the persistence threshold. The blue circles and black dots represent the clusters and maximum-density critical points in a simulation box, respectively. The left box shows the case of a low $\sigma$ parameter choice, where there is a large number of CP$_{\rm max}$ present and a higher fraction of clusters contain a CP$_{\rm max}$. Hence, the completeness is high and the purity is low in this case. The right box shows the case of a high $\sigma$ parameter choice, resulting in less CP$_{\rm max}$. This leads to low completeness and high purity. 

\begin{figure}[t]
    \centering
    \includegraphics[width=0.49\linewidth]{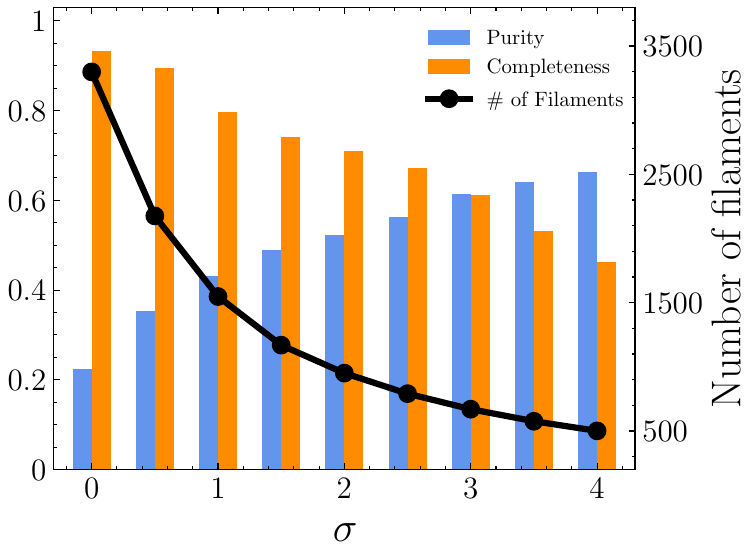}\hspace{5pt}
        \includegraphics[width=0.45\linewidth]{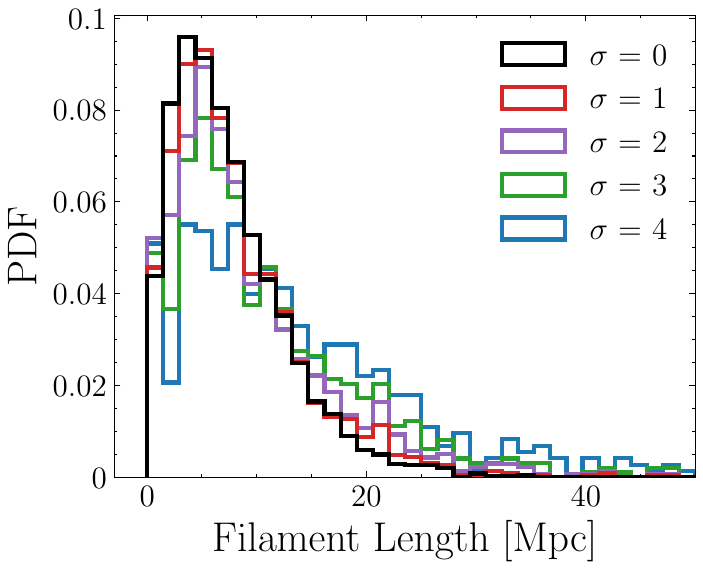}
    \caption{Left: Completeness (orange bars), purity (blue bars) and the number of identified filaments (black circles) for various $\sigma$ values for the persistence threshold. The persistence threshold is calibrated to simultaneously maximize the completeness and purity of the reconstructed filamentary structure. Right: Distribution of filament lengths for various persistence thresholds. 
    }
    \label{calibration}
\end{figure}

Figure~\ref{calibration} shows the results of our calibration of the $\sigma$ parameter. 
The left panel shows the completeness and purity of the reconstructed filamentary structure for different persistence thresholds, along with the corresponding number of filaments identified. It is clear that completeness and purity are simultaneously maximized for $\sigma=3$.
We therefore utilize the catalogue of filaments that is generated using this persistence threshold in the following analyses. The right panel of the figure shows the distribution of filament lengths for various persistence thresholds. As seen in the right panel, the number of shorter filaments decreases as $\sigma$ increases. This is because a higher persistence threshold reduces the number of spurious filaments, which are generally shorter.  

In DisPerSE, filaments are defined by a series of connected segments branching between critical points. In order to avoid contamination from the high density endpoints of filaments, we remove any filamentary segments that lie within  the $3\times R_{\rm 200}$ radius\footnote{We have checked that our results do not significantly change if  we remove filamentary segments within $5\times R_{\rm 200}$.} of any clusters which contain a CP$_{\rm max}$ \cite{Galarraga-Espinosa:2021kbe}. This ensures that we are only considering the body of the filament itself and not the higher density end points. Additionally, in order to avoid an underestimation of filament densities near the edges of the simulation box, we do not consider any filament segments which are within 10~Mpc of the box edge. We therefore effectively only consider filaments whose spines lie within the (90~Mpc)$^3$ sub-volume of the total simulation volume. This ensures that all DM particles within a radial distance of 10 Mpc from each filament's spine are included in the analysis. Before applying these cuts, our catalogue contains 679 filaments. Once the cuts have been applied, our final filament catalogue contains 269 filaments with minimum and maximum lengths of 0.15 and 48.3 Mpc, respectively, along with an average length of 9.13 Mpc and a median length of 6.63 Mpc. We discuss the radii of the filaments in our catalogue in section~\ref{sec:density}.

\section{Filament density profiles}
\label{sec:density}
\begin{figure}[t]
    \centering
    \includegraphics[width=0.5\linewidth]{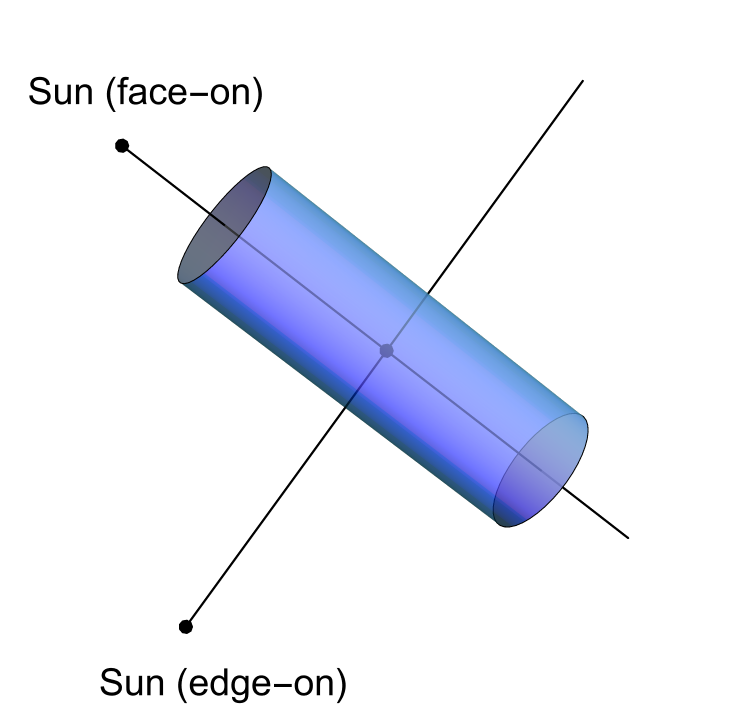}
    \caption{The two considered orientations of the Sun and the observer with respect to a filament: an edge-on view and a face-on view. 
    }
    \label{fig:filaments}
\end{figure}

The DM decay signals from filaments depend on the DM density profile within these cosmic structures. To simplify the calculation of the DM decay signals, we assume a cylindrical geometry and consider two orientations for the Sun and the observer with respect to the filaments: an edge-on view perpendicular to the spine of the filament, and a face-on view along the spine of the filament, as shown in figure~\ref{fig:filaments}. For the case of the edge-on view, the radial DM density profile, computed as a function of the cylindrical radial distance from the spine of the filament, enters the analysis. For the face-on view, the DM density profile as a function of the length of the filament is needed. 
As we will discuss in section~\ref{sec:DM}, the DM column density (i.e. the integral over the line of sight of the DM density profile) enters the calculation of the DM decay signals. This integral has the smallest value in the edge-on view and the largest in the face-on view. In general, the line of sight could go through the filament at an arbitrary angle with respect to the filament's spine.  
We consider the edge-on and face-on views as the lower and upper limits to the general result, respectively.

\begin{figure}[t]
    \centering
    \includegraphics[width=0.48\linewidth]{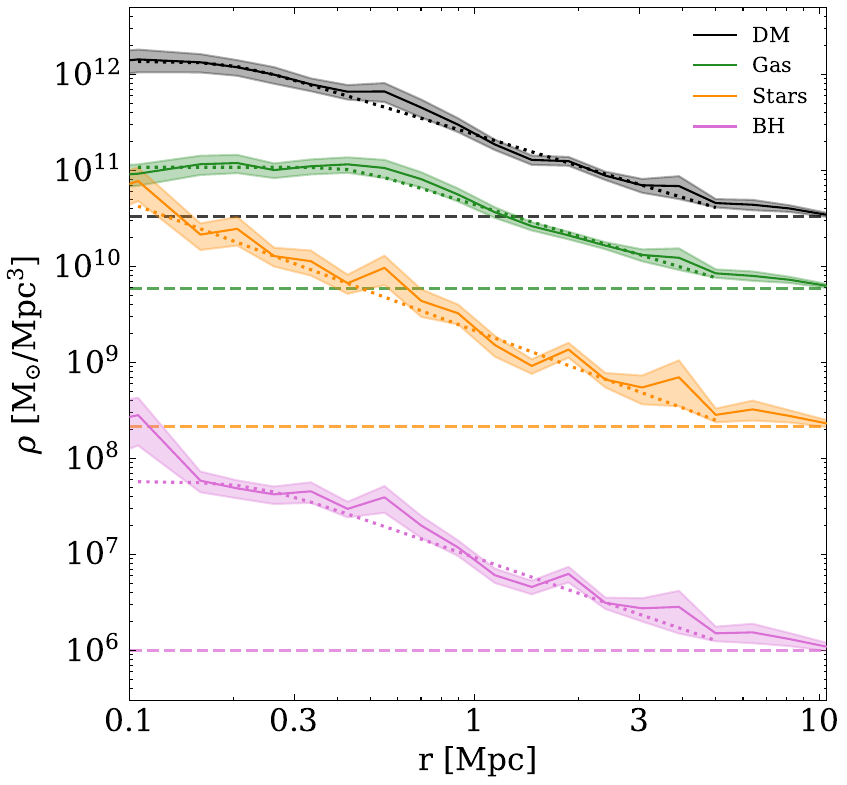}\hspace{5pt}
        \includegraphics[width=0.48\linewidth]{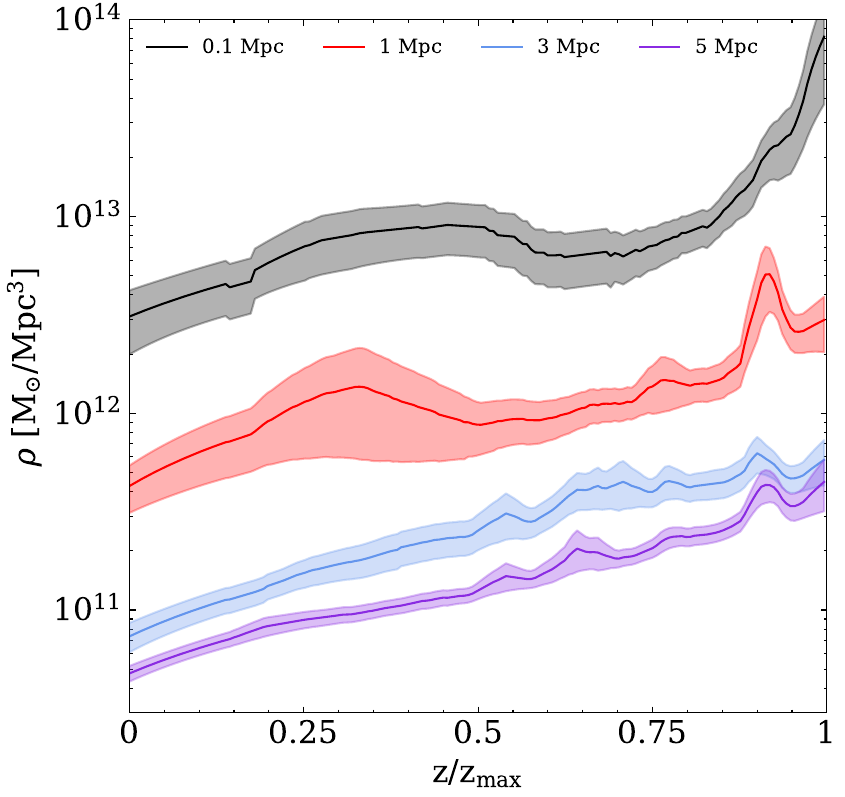}
    \caption{Left: The average radial density profiles of DM (black), gas (green), stars (orange) and black holes (pink) for all filaments. The solid colored lines indicate the mean density of the filaments, and the shaded bands correspond to the standard error of the mean density. The dotted lines correspond to the best fit $\beta$-model (eq.~\eqref{eq:betamodel}) to the average density profiles from the simulations. The fit parameters and reduced $\chi^2$ values are shown in table \ref{tab:fitvals}. The dashed colored lines correspond to the mean density of each matter component in the full simulation volume.  
    Right: The average DM density profiles (solid lines) and their standard error (shaded bands) for all filaments along the length of the filaments, for radial distances of 0.1, 1, 3 and 5 Mpc from the spine. The distance along the length of each filament, $z$,  has been normalized by the total length of the filament, $z_{\rm max}$ such that $z/z_{\text{max}}=0$ corresponds to the low density end of the filament defined by a saddle critical point and $z/z_{\text{max}}=1$ corresponds to the high density end of the filaments defined by a CP$_{\textrm{max}}$.}
    \label{dens_profiles}
\end{figure}

The left panel of figure~\ref{dens_profiles} shows the average density profile of all filaments for different matter components (DM, stars, gas and black holes) within the filaments, as a function of the cylindrical radial distance from the filament spines. The density profiles are first calculated for each  
segment composing each filament, which are then averaged to find the density profile of the entire filament. 
Then, the average of all filament density profiles is computed to obtain the final density profile, specified by the solid colored lines in the figure. The shaded bands correspond to the standard error of the mean density profile. For comparison, the dashed colored lines show the mean density in the simulation box for each matter component. The figure shows that the mean density of all matter components in the filaments reach their respective mean box densities at a radial distance of $\sim 10$~Mpc from the filament spines. 

The right panel of figure~\ref{dens_profiles} shows the average DM density profile of all filaments along the lengths of the filaments at various radial distances from the spine. The solid lines and shaded bands correspond to the average DM density and its standard error, respectively. The distance along the length of each filament, $z$, has been normalized by the filament's total length, $z_{\rm max}$, such that $z/z_{\rm max}=0$ corresponds to the saddle critical point and $z/z_{\rm max}=1$ corresponds to the CP$_{\textrm{max}}$. 
The plot clearly shows an expected increase in density along the length of filaments as one goes from the saddle to the maximum critical point defining each filament. Additionally, the overall amplitude of the curves decreases with increasing radial distance, as expected.

Historically, the radius of a cosmic filament has not been rigorously defined. However, a recent study~\cite{Wang:2024qej} proposed that the radius at which the slope of the radial galaxy number density profile becomes minimum is a physically motivated definition for the filament radius. We compute the logarithmic slope of the mean DM density profile, shown in the left panel of figure~\ref{dens_profiles}, and find that the slope reaches a minimum at a radius of $\sim 1$~Mpc, which is consistent with the findings of ref.~\cite{Wang:2024qej}. This suggests that the average radius of the filaments in our catalogue is $\sim 1$~Mpc. Alternatively, we can determine the radius at which the DM density profile in a filament reaches a factor of a few times the mean DM density in the simulation box (dashed black line in the left panel of figure~\ref{dens_profiles}), and consider that as the radius of the filament. For example, if we choose the radius of the filament to be the radial distance where the DM density is twice the mean density in the simulation box, then the average radius of filaments is 2.5 Mpc for filaments with length $<$ 6 Mpc and 1.3 Mpc for filaments with length $>$ 6 Mpc. The average radius of all filaments, defined in this way, is $\sim$ 2 Mpc. Using this definition, we find that 4.2\% of the volume of the Universe is contained within cosmic filaments. Note also that, in general, shorter filaments tend to have larger radii. This reflects the environmental differences between short and long filaments. Short filaments are typically located near clusters, bridging small gaps between them, while long filaments extend across underdense regions, connecting larger areas \cite{Galarraga-Espinosa:2021kbe}.

In our analysis, we directly use the DM density profiles extracted from the simulations to compute the DM decay signals. However, for completeness we also investigate how well the average radial density profile of different matter components in the filaments can be fitted with a fitting function proposed in the literature and adopted in previous works~\cite{Cavaliere:1976tx, Ettori:2013tka, Galarraga-Espinosa:2021kbe}. In particular, we consider the $\beta$-model  given by
\begin{equation}
    \rho(r) = \frac{\rho_0}{\left[ 1 + \left(\frac{r}{r_0} \right)^\alpha \right]^\beta} \; ,
    \label{eq:betamodel}
\end{equation}
with four free parameters $\rho_0$, $r_0$, $\alpha$, and $\beta$.

We obtain the best-fit $\beta$-model to the average density profiles of all the filaments from the simulations by minimizing the $\chi^2$ function,
\begin{equation}
    \chi^2 = \sum_i^N \frac{(y_i - \rho(r_i, \textbf{p}))^2}{\sigma_i^2} \; ,
    \label{eq:chi2_}
\end{equation}
where $N$ is the number of data points in the density profile, $y_i$ is  the value of the average density at the cylindrical radial distance $r_i$ from the filament spine, $\sigma_i$ is the corresponding standard error of the density,  and $\textbf{p}$ are the four free parameters determined by the minimization procedure.

Table~\ref{tab:fitvals} lists the best-fit parameters of the $\beta$-model, along with the reduced-$\chi^2$ values for the DM, gas, stars, and black holes in the filaments. The reduced-$\chi^2$ is defined as $\bar{\chi}^2$ $\equiv$ $\chi^2/(N-{\rm dof})$, where ${\rm dof}=4$ is the number of fitting parameters. The $\bar{\chi}^2$ values indicate an excellent fit for the DM density profile.

We also investigated whether the Navarro-Frenk-White (NFW)~\cite{Navarro:1996gj} profile, which provides a good fit to the DM density profiles in galaxies, could similarly fit the DM density profiles of filaments. We find that the NFW profile provides a less accurate fit to the density profile of filaments for all matter components compared to the $\beta$-model, with $\bar{\chi}^2$ values of 1.3, 10.6, 1.9, and 1.6 for the DM, gas, stars, and black hole density profiles, respectively. This result is expected, as the NFW profile, with only two free parameters, is not well-suited to describe filament density profiles, especially for the DM and gas components at small radial distances, where their profiles flatten. 

\begin{table*}[t]
	\centering
	\begin{tabular}{|c|c|c|c|c|c|} 
            \hline
		 \text{ } & $\rho_0$ [Mpc/M$_{\odot}$] &  r$_0$ [Mpc] & $\alpha$ & $\beta$& $\bar{\chi}^2$\\
				\hline
    Dark matter & 1.4$\times$10$^{12}$ & 0.20 & 6.81 & 0.16 & 0.86\\
    Gas & 1.1$\times$10$^{11}$ & 0.44 & 10.9 & 0.10 & 0.62 \\
    Stars & 9.2$\times$10$^{10}$ & 0.06 & 4.95 & 0.27 &  1.44\\
    Black holes & 5.7$\times$10$^{7}$ & 0.23 & 5.63 & 0.22 &  1.54\\
    \hline
	\end{tabular}
	\caption{Best-fit parameters of the $\beta$-model (eq.~\eqref{eq:betamodel}) and the reduced-$\chi^2$ values for each of the four matter components, obtained by fitting the respective average radial density profiles of the simulated filaments with the $\beta$-model.
    }
 \label{tab:fitvals}
\end{table*}

\section{Dark matter signal}
\label{sec:DM}
Dark matter particles inside filaments could decay into electron-positron pairs. These secondary particles can produce synchrotron radiation in the presence of filamentary magnetic fields. The higher the DM density, the higher the number of electrons, and the larger the final synchrotron photon flux.
The electron flux produced by the DM decays is given by
\begin{equation}
S_{e}(E_e)= \frac{\Sigma_\text{DM}}{\tau_D m_{\chi} c^2} \, \frac{d N_{e}}{d E_e}(E_e) \; ,
\end{equation}
where $E_e$ is the energy of an electron produced in the decay, $\tau_D$ is the DM particle decay lifetime, $m_{\rm DM}$ is the DM mass, $\de N_{e}/ \de E_e$ is the electron spectrum produced in the decay event, and $\Sigma_\text{DM}$ is the DM column density. Given the filament DM density profiles extracted from the simulations and shown in figure~\ref{dens_profiles}, we find the DM column densities in case of the face-on and edge-on orientations, 
\begin{align}
\Sigma^\textrm{face-on}_\textrm{DM} =& \text{ }(8.15_{-2.80}^{+2.82})\times10^{13}~ \Msun \,\textrm{Mpc}^{-2}\; , \nonumber\\
\Sigma^\textrm{edge-on}_\textrm{DM} =& \text{ }(2.36_{-0.39}^{+0.39})\times10^{12}~ \Msun \,\textrm{Mpc}^{-2} \; .
\end{align}

The face-on DM column density is obtained by integrating the mean DM density profile at a radial distance of 0.1~Mpc along a filament length of 8 Mpc. This length is chosen because it lies between the average and median filament lengths. To find the edge-on DM column density, we integrate the mean DM density profile over the radial distance from 0.1 to 10~Mpc, measured from the spine of the filaments. To be conservative, we take a lower integration limit of 0.1~Mpc, since below this radial distance, the position of the filament spines identified in DisPerSE is uncertain~\cite{Galarraga-Espinosa:2020rhp, Galarraga-Espinosa:2021kbe}. The upper integration limit is set to 10~Mpc, corresponding to the radial distance at which the filament's DM density profile reaches the mean DM density in the simulation volume. Note that we do not use the filament radius as the upper limit,
in order to capture the full DM density profile of the filaments in the edge-on column density calculation.
 The upper and lower uncertainties on the DM column densities are obtained by integrating the upper and lower uncertainty bands around the mean DM density profiles of the filaments, respectively. 
Following the treatment of ref.~\cite{Vernstrom:2021hru}, the electron differential number density reads
\begin{equation}
n_{e}(E_e)=\frac{1}{b_\textrm{tot}(E_e)} \int_{E_e}^{\infty} \de E_e^{\prime} \, S_{e}(E_e^{\prime}) \; ,
\end{equation}
where $b_\textrm{tot} = b_{\text {sync}} + b_{\text {ICS}}$ is the total energy loss due to synchtron radiation and inverse-Compton scattering on the cosmic microwave background. These energy losses can be expressed as $b_i = 4 \sigma_{\text{th}} \,  E_e^2 \, u_i / (3 \, m_e^2)$ with $i = \{\text{sync}, \text {ICS} \}$, $\sigma_{\text{th}}$ is the Thomson cross-section, $u_{\text{CMB}} = 0.260 \; \text{eV}/\text{cm}^3$ is the energy density of the cosmic microwave background, and $u_B = B^2 / 2 \mu_0$ is the magnetic energy density, with $B$ denoting the magnetic field strength and $\mu_0$ being the vacuum permeability.\footnote{Energy losses due to synchrotron radiation are subdominant for the typical magnetic fields assumed inside filaments. Also, ref. \cite{Vernstrom:2021hru} shows that the confinement time is typically larger than the cooling time, therefore, we neglect the diffusion of electrons in this work.} 
The power emitted per unit frequency by a single relativistic electron of energy $E_e$ moving in a magnetic field is described by the radiative synchrotron radiated power as \cite{longair2011high}
\begin{equation}
P_{\text {sync}}(E_e, \nu)=\frac{\sqrt{3} e^{3} c}{4\pi\epsilon_0 \, m_{e} c^2} B F\left(\nu / \nu_{c}\right) \; ,
\label{eq:power}
\end{equation}
where $\nu_{c} \equiv 3 c^2 e /(4 \pi) /\left(m_{e} c^{2}\right)^{3} B \, E_e^{2}$ is the critical frequency. The synchrotron radiation spectrum reads $F(t) \equiv t \int_{t}^{\infty} \de z \, K_{5 / 3}(z)$, where $K_{5 / 3}$ is the modified Bessel function of the second kind of order 5/3.
The synchrotron emissivity is obtained as the convolution of the electron number density and the radiative synchrotron power via \begin{equation}
j_{\text {sync}}(\nu) = \int \de E_e \, P_{\text {sync}}(E_e, \nu) \, n_{e}(E_e) \; , 
\end{equation}
and the intensity at frequency $\nu$ is related to the emissivity via
\begin{equation}
I(\nu) = \frac{1}{\Delta\Omega}\int \frac{\de \Omega }{4\pi}\, j_{\rm sync}(\nu)  \; ,
\end{equation}
where $\Omega$ is the solid angle that the filament subtends, and $\Delta \Omega$ specifies the field-of-view to the filaments~\cite{Vernstrom:2021hru}.  In the next section, we present our results in terms of the 
brightness temperature, defined as
\begin{equation}
    T_{b}= \frac{I(\nu) \, c^{2}}{2 k_B \nu^{2}} \; ,
\end{equation}
 where $k_B$ is the Boltzmann constant. In this context, the brightness temperature quantifies the intensity of radiation from DM decays in filaments, expressed in temperature units.

\section{Results}
\label{sec:results}
The formalism introduced in section~\ref{sec:DM} is used to estimate the expected brightness temperature from a DM-induced radio signal over a range of frequencies. For DM masses in the  1--100~GeV range and magnetic fields typical of filamentary structures (on the order of a few to hundreds of nanoGauss), the DM-induced synchrotron radiation will fall in the radio band. This is especially interesting in view of the first robust detection of a stacked radio signal from large inter-cluster filaments reported by ref.~\cite{Vernstrom:2021hru}. The authors showed that the measured radio signal is 30--40 times larger than predicted by simulations of the shocked intergalactic gas, for the four radio maps considered in their analysis (154 MHz, 118 MHz, 88 MHz, 73 MHz).\footnote{Ref.~\cite{Vernstrom:2021hru} used observations from the GaLactic and Extragalactic All-sky Murchison Widefield Array (GLEAM) survey \cite{Wayth15,Hurley-Walker17} and the Owens Valley Radio Observatory Long Wavelength Array (OVRO-LWA) survey \cite{Eastwood18}. } The frequency of the four maps and the observed brightness temperature with the corresponding standard deviation is listed in the first and second columns of table~\ref{tab:Tb}, respectively.  
 
This paper explores the possibility that this radio excess could be explained in terms of DM particles decaying into electrons, subsequently emitting synchrotron radiation. The expected brightness temperature for a DM-induced radio signal is shown in table~\ref{tab:Tb} for the benchmark case of a DM particle with $m_\text{DM} =30$~GeV, decay lifetime $\tau = 2 \times 10^{27}$~s and filamentary magnetic field $B=30$~nG. The density profile is assumed to be the mean between the face-on and edge-on orientations. We find that for all four frequencies under consideration, the observed brightness temperature can be explained by a DM-induced signal in a region of the DM parameter space not yet excluded by other observations. The uncertainties are discussed below.

\begin{table}[t]
    \centering
    \renewcommand{\arraystretch}{1.5} 
    \begin{tabular}{|c|c|c|c|}
        \hline
        \rule{0pt}{12pt} \textbf{$\nu$} [MHz] & T$_b^\textrm{obs}$ [K]  & T$_b^\textrm{DM}$ [K] \\ 
        \hline
        154 & $0.10 \pm 0.04$ & $0.10$ \\  
        118 & $0.22 \pm 0.06$ & $0.22$  \\  
        88 & $0.44 \pm 0.09 $ & $0.51$  \\  
        73 & $1.1 \pm 0.2$ & $0.85$ \\  
        \hline
    \end{tabular}
    \caption{Comparison between the observed brightness temperature (second column) of a filament in different frequencies (first column) and the expected brightness temperature associated with a DM signal (third column), for a DM particle with $m_\text{DM} =30$~GeV, decay lifetime $\tau = 2 \times 10^{27}$~s and filamentary magnetic field $B=30$~nG. See the text for details.}
    \label{tab:Tb}
\end{table}

If we assume that the observed signal is due to astrophysical backgrounds instead of DM particles, then we can use the non-observation of a DM signal to draw bounds on the DM decay lifetime. Following refs.~\cite{Cowan:2010js, Janish:2023kvi, Pinetti:2025owq}, we define the goodness-of-fit $\chi^2$,

\begin{equation} 
     \chi^2 (\tau) = \sum_{i} \left(\frac{T_b^\textrm{DM}(\nu_{i} | \tau) - d_{i}}{\sigma_{i}} \right)^2 \; ,
     \label{eq:chi2}
\end{equation}
where $d_i$ represents the observed $T_b^\textrm{obs}$, $\sigma_{i}$ its error, and $T_b^\textrm{DM}$ the expected DM-induced signal ($T_b^\textrm{Face-on}$ or $T_b^\textrm{Edge-on}$, depending on the configuration). The index $i$ runs over the four observed frequencies.  
 We determine the constraints on the DM decay lifetime, $\tau$, by evaluating the $\chi^2 (\tau)$ statistics defined above while scanning over $\tau$. The best-fit value of $\tau$ corresponds to the minimum of $\chi^2 (\tau)$, while the 95\% confidence level is defined as the range of $\tau$ for which $\Delta \chi^2 \leq 3.84$ relative to the global minimum.

\begin{figure}[t]
    \centering
    \includegraphics[width=0.495\linewidth]{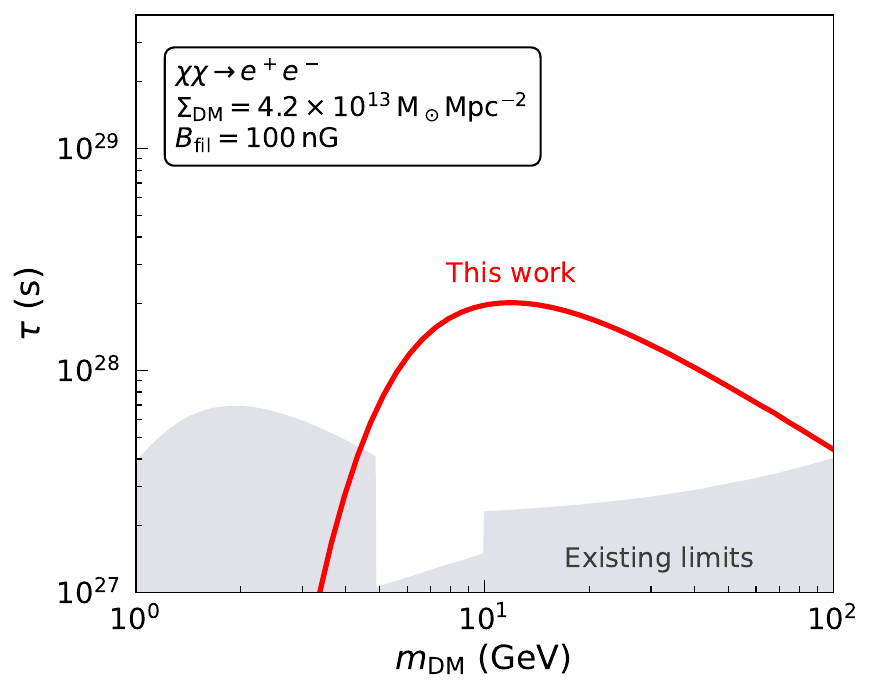}
    \includegraphics[width=0.495\linewidth]{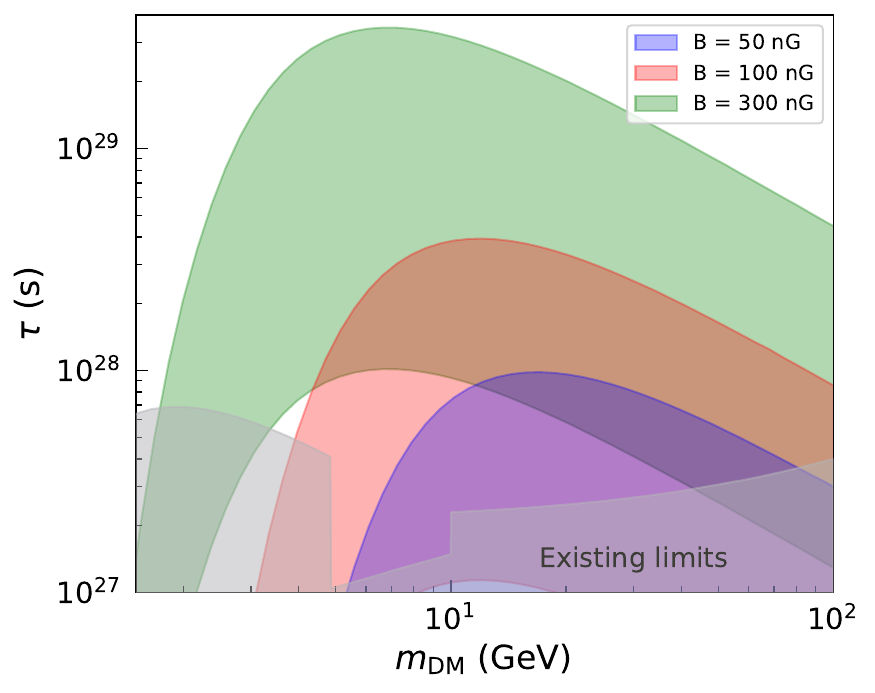}
    \caption{Lower limits at 95\% confidence level on the DM decay lifetime, $\tau$, as a function of the DM mass, $m_\textrm{DM}$. 
    The red curve in the left panel shows the limits assuming a filament magnetic field of 100~nG, with the DM column density taken as the average between the face-on and edge-on values. In the right panel, the limits are shown for three different values of the filament magnetic field: 50~nG (blue), 100~nG (red), and 300~nG (green). The shaded error bands represent the uncertainty associated with the DM density profile in filaments. Existing limits are shown in gray in both panels.}
    \label{fig:bounds}
\end{figure}
Intercluster filaments act as an irreducible source of magnetic fields with present-day average field strengths ranging from a few nG up to 600 nG~\cite{Vernstrom:2021hru, Brown:2017dwx, Vacca:2018rta, OSullivan:2018shr, Vernstrom:2019gjr, OSullivan:2020pll, Locatelli:2021byc, Amaral:2021mly, Carretti:2022tbj, Hoang:2023uyx, Anderson:2024giz, Carretti:2024bcf}. 
The 95\% confidence-level lower limits on the DM decay lifetime, $\tau$, as a function of the DM mass, $m_{\rm DM}$, are presented in the left panel of
figure~\ref{fig:bounds} for a fiducial magnetic field strength of 100 nG, and taking the DM density profile as the mean between the face-on and edge-on orientations (red curve). The grey area is already excluded by X-ray and gamma-ray observations~\citep{Cirelli:2020bpc, Cirelli:2023tnx, Essig:2013goa, Massari:2015xea, Blanco:2018esa}, charged cosmic ray measurements~\citep{Fornengo:2013xda, Giesen:2015ufa, Boudaud:2016mos}, cosmic microwave background and Lyman-$\alpha$ forest~\citep{Slatyer:2016qyl, Capozzi:2023xie, Xu:2024vdn, Liu:2020wqz},  and heating of gas-rich dwarf galaxies~\citep{Wadekar:2021qae}. We also note that the bounds derived in this work are more than two orders of magnitude stronger than those obtained under the assumption of a constant DM density in filaments, as adopted in ref.~\cite{Vernstrom:2021hru}.

The right panel of figure~\ref{fig:bounds} accounts for the uncertainties on the constraints due to the spatial orientation and
choice of the filamentary magnetic field.  The constraints are presented for three different choices of the filamentary magnetic fields: 50~nG (blue), 100~nG (red), and 300~nG (green). The upper limit of each band refers to the face-on orientation, while the lower limit is associated with the edge-on configuration. We find that the constraints derived in this analysis are competitive for magnetic fields above 30~nG and improve by over two orders of magnitude compared to the state-of-the-art in the literature for certain masses in the case of strong filamentary magnetic fields ($\gtrsim 300$~nG).

\section{Discussion and conclusions}
\label{sec:conclusions}
Cosmic filaments form the structure of the cosmic web, yet many aspects of their properties and evolution remain uncertain. Understanding the large-scale structure of the Universe requires an accurate characterization of both dark matter (DM) and baryonic matter within these cosmic bridges.

In this work, we leverage the EAGLE hydrodynamical simulations to investigate the DM density profiles of filaments and their implications for DM decay signals. Using the DisPerSE code, we identify 269 filaments in the simulations, spanning lengths from $0.15$ to $48.3$~Mpc with an average of $9.13$~Mpc. We extract the DM density profiles as a function of radial distance from the filament spine and explore their dependence on the filament length. Additionally, we compute the expected DM decay signals from filaments, considering two distinct observer orientations: an edge-on view, perpendicular to the filament spine, and a face-on view, aligned along the spine of the filament.

We find that DM particles with GeV-scale masses, decaying into electron-positron pairs, can produce detectable radio emission via synchrotron radiation. The predicted DM-induced signal is consistent with the radio excess reported in ref.~\cite{Vernstrom:2021hru}. If this excess instead originates from astrophysical backgrounds, we use it to derive constraints on the DM decay lifetime. For typical filamentary magnetic field strengths, our results provide competitive bounds in the mass range of 1--100~GeV, improving upon existing constraints by a factor of a few in the case of weak magnetic fields and by up to two orders of magnitude for strong filamentary magnetic fields.

These findings motivate a systematic investigation of astrophysical backgrounds, as well as a comprehensive effort to better characterize the physical properties of filaments, particularly their DM density profiles and magnetic fields.
Future studies could explore the role of subhaloes within filaments and their potential impact on signals from annihilating DM. The effect of alternative cosmological scenarios beyond the standard  $\Lambda$CDM paradigm—including warm DM and self-interacting DM—remains an open question. Moreover, the EAGLE simulations used in this study do not include magnetohydrodynamics. Future analyses leveraging large-volume magnetohydrodynamical simulations will be crucial in unraveling the evolution of magnetic fields within filaments. The DM distribution in filaments is also sensitive to the AGN feedback considered in the simulations (see e.g.~\cite{Hadzhiyska:2024qsl}). Comparing the results of different cosmological simulations with different AGN feedback would provide important insight. We leave these exciting directions for future work.

Finally, upcoming multi-wavelength observations will offer unprecedented opportunities to probe the nature of cosmic filaments. Infrared studies with JWST, Roman, Euclid, and SPHEREx, X-ray observations with XRISM, ATHENA, and eROSITA, and optical surveys with LSST will collectively pave the way for groundbreaking discoveries of these fundamental components of the cosmic web.

\acknowledgments
We would like to thank Aurelio Amerio, Cyril Creque-Sarbinowski, Chris Dessert, and David Dunsky for helpful discussion. For facilitating portions of this research, EP and NB acknowledge the Center for Theoretical Underground Physics and Related Areas (CETUP*), The Institute for Underground Science at Sanford Underground Research Facility (SURF), and the South Dakota Science and Technology Authority for hospitality and financial support, as well as for providing a stimulating environment during the CETUP* 2023 Workshop. EP is grateful for the hospitality of Perimeter Institute where part of this work was carried out. Research at Perimeter Institute is supported in part by the Government of Canada through the Department of Innovation, Science and Economic Development and by the Province of Ontario through the Ministry of Colleges and Universities. This work was also supported by a grant from the Simons Foundation (1034867, Dittrich). EV and NB acknowledge the support of the Natural Sciences and Engineering Research Council of Canada (NSERC), funding reference number RGPIN-2020-07138 and the NSERC Discovery Launch Supplement, DGECR-2020-00231. NB acknowledges the support of the Canada Research Chairs Program. We have used simulations from the EAGLE Project public data release~\citep{EAGLE:2017} available at http://icc.dur.ac.uk/Eagle/database.php.

\typeout{}
\bibliographystyle{JHEP}
\bibliography{refs}

\providecommand{\href}[2]{#2}\begingroup\raggedright\begin{thebibliography}{100}

\bibitem{Bond:1995yt}
J.~R. Bond, L.~Kofman, and D.~Pogosyan, {\it {How filaments are woven into the
  cosmic web}},  {\em Nature} {\bf 380} (1996) 603--606,
  [\href{http://arxiv.org/abs/astro-ph/9512141}{{\tt astro-ph/9512141}}].

\bibitem{White:1987yr}
S.~D.~M. White, C.~S. Frenk, M.~Davis, and G.~Efstathiou, {\it {Clusters,
  filaments, and voids in a universe dominated by cold dark matter}},  {\em
  Astrophys. J.} {\bf 313} (1987) 505--516.

\bibitem{Peebles:1980yev}
P.~J. Peebles, {\em {The Large-Scale Structure of the Universe}}.
\newblock Princeton University Press, 11, 1980.

\bibitem{Zeldovich:1982zz}
Y.~B. Zeldovich, J.~Einasto, and S.~F. Shandarin, {\it {Giant Voids in the
  Universe}},  {\em Nature} {\bf 300} (1982) 407--413.

\bibitem{Bahcall:1988ch}
N.~A. Bahcall, {\it {Large-scale structure in the universe indicated by galaxy
  clusters}},  {\em Ann. Rev. Astron. Astrophys.} {\bf 26} (1988) 631--686.

\bibitem{Gott:1988rj}
J.~R. Gott, III et~al., {\it {THE TOPOLOGY OF LARGE SCALE STRUCTURE. 3.
  ANALYSIS OF OBSERVATIONS}}, .

\bibitem{Margon:1998vu}
B.~Margon, {\it {The Sloan Digital Sky Survey}},  {\em Phil. Trans. Roy. Soc.
  Lond. A} {\bf 357} (1999) 93--103,
  [\href{http://arxiv.org/abs/astro-ph/9805314}{{\tt astro-ph/9805314}}].

\bibitem{2dFGRS:2005yhx}
{\bf 2dFGRS} Collaboration, S.~Cole et~al., {\it {The 2dF Galaxy Redshift
  Survey: Power-spectrum analysis of the final dataset and cosmological
  implications}},  {\em Mon. Not. Roy. Astron. Soc.} {\bf 362} (2005) 505--534,
  [\href{http://arxiv.org/abs/astro-ph/0501174}{{\tt astro-ph/0501174}}].

\bibitem{Dubois:2014lxa}
Y.~Dubois et~al., {\it {Dancing in the dark: galactic properties trace spin
  swings along the cosmic web}},  {\em Mon. Not. Roy. Astron. Soc.} {\bf 444}
  (2014), no.~2 1453--1468, [\href{http://arxiv.org/abs/1402.1165}{{\tt
  arXiv:1402.1165}}].

\bibitem{Chen:2015oqa}
Y.-C. Chen, S.~Ho, R.~Mandelbaum, N.~A. Bahcall, J.~R. Brownstein, P.~E.
  Freeman, C.~R. Genovese, D.~P. Schneider, and L.~Wasserman, {\it {Detecting
  Effects of Filaments on Galaxy Properties in the Sloan Digital Sky Survey
  III}},  {\em Mon. Not. Roy. Astron. Soc.} {\bf 466} (2017), no.~2 1880--1893,
  [\href{http://arxiv.org/abs/1509.06376}{{\tt arXiv:1509.06376}}].

\bibitem{Kuutma:2017yvb}
T.~Kuutma, A.~Tamm, and E.~Tempel, {\it {From voids to filaments: environmental
  transformations of galaxies in the SDSS}},  {\em Astron. Astrophys.} {\bf
  600} (2017) L6, [\href{http://arxiv.org/abs/1703.04338}{{\tt
  arXiv:1703.04338}}].

\bibitem{Laigle:2017byb}
C.~Laigle et~al., {\it {COSMOS2015 photometric redshifts probe the impact of
  filaments on galaxy properties}},  {\em Mon. Not. Roy. Astron. Soc.} {\bf
  474} (2018), no.~4 5437--5458, [\href{http://arxiv.org/abs/1702.08810}{{\tt
  arXiv:1702.08810}}].

\bibitem{Sarron:2019cyf}
F.~Sarron, C.~Adami, F.~Durret, and C.~Laigle, {\it {Pre-processing of galaxies
  in cosmic filaments around AMASCFI clusters in the CFHTLS}},  {\em Astron.
  Astrophys.} {\bf 632} (2019) A49,
  [\href{http://arxiv.org/abs/1903.02879}{{\tt arXiv:1903.02879}}].

\bibitem{Bonjean:2019hnz}
V.~Bonjean, N.~Aghanim, M.~Douspis, N.~Malavasi, and H.~Tanimura, {\it
  {Filament profiles from WISExSCOS galaxies as probes of the impact of
  environmental effects}},  {\em Astron. Astrophys.} {\bf 638} (2020) A75,
  [\href{http://arxiv.org/abs/1912.06559}{{\tt arXiv:1912.06559}}].

\bibitem{Bonjean:2017okn}
V.~Bonjean, N.~Aghanim, P.~Salom\'e, M.~Douspis, and A.~Beelen, {\it {Gas and
  galaxies in filaments between clusters of galaxies - The study of
  A399-A401}},  {\em Astron. Astrophys.} {\bf 609} (2018) A49,
  [\href{http://arxiv.org/abs/1710.08699}{{\tt arXiv:1710.08699}}].

\bibitem{Tanimura:2019uxm}
H.~Tanimura, N.~Aghanim, V.~Bonjean, N.~Malavasi, and M.~Douspis, {\it {Density
  and temperature of cosmic-web filaments on scales of tens of megaparsecs}},
  {\em Astron. Astrophys.} {\bf 637} (2020) A41,
  [\href{http://arxiv.org/abs/1911.09706}{{\tt arXiv:1911.09706}}].

\bibitem{Vernstrom:2021hru}
T.~Vernstrom, G.~Heald, F.~Vazza, T.~J. Galvin, J.~West, N.~Locatelli,
  N.~Fornengo, and E.~Pinetti, {\it {Discovery of magnetic fields along stacked
  cosmic filaments as revealed by radio and X-ray emission}},  {\em Mon. Not.
  Roy. Astron. Soc.} {\bf 505} (2021), no.~3 4178--4196,
  [\href{http://arxiv.org/abs/2101.09331}{{\tt arXiv:2101.09331}}].

\bibitem{Planck:2018vyg}
{\bf Planck} Collaboration, N.~Aghanim et~al., {\it {Planck 2018 results. VI.
  Cosmological parameters}},  {\em Astron. Astrophys.} {\bf 641} (2020) A6,
  [\href{http://arxiv.org/abs/1807.06209}{{\tt arXiv:1807.06209}}]. [Erratum:
  Astron.Astrophys. 652, C4 (2021)].

\bibitem{Dunsky:2025pvd}
D.~I. Dunsky, G.~Krnjaic, and E.~Pinetti, {\it {Observing Dark Matter Decays to
  Gravitons via Graviton-Photon Conversion}},
  \href{http://arxiv.org/abs/2503.19019}{{\tt arXiv:2503.19019}}.

\bibitem{EAGLE:2017}
{The EAGLE team}, {\it {The EAGLE simulations of galaxy formation: Public
  release of particle data}},  {\em arXiv e-prints} (June, 2017)
  arXiv:1706.09899, [\href{http://arxiv.org/abs/1706.09899}{{\tt
  arXiv:1706.09899}}].

\bibitem{Schaye:2014tpa}
J.~Schaye et~al., {\it {The EAGLE project: Simulating the evolution and
  assembly of galaxies and their environments}},  {\em Mon. Not. Roy. Astron.
  Soc.} {\bf 446} (2015) 521--554, [\href{http://arxiv.org/abs/1407.7040}{{\tt
  arXiv:1407.7040}}].

\bibitem{Crain:2015poa}
R.~A. Crain et~al., {\it {The EAGLE simulations of galaxy formation:
  calibration of subgrid physics and model variations}},  {\em Mon. Not. Roy.
  Astron. Soc.} {\bf 450} (2015), no.~2 1937--1961,
  [\href{http://arxiv.org/abs/1501.01311}{{\tt arXiv:1501.01311}}].

\bibitem{McCarthy:2016mry}
I.~G. McCarthy, J.~Schaye, S.~Bird, and A.~M.~C. Le~Brun, {\it {The BAHAMAS
  project: Calibrated hydrodynamical simulations for large-scale structure
  cosmology}},  {\em Mon. Not. Roy. Astron. Soc.} {\bf 465} (2017), no.~3
  2936--2965, [\href{http://arxiv.org/abs/1603.02702}{{\tt arXiv:1603.02702}}].

\bibitem{2018MNRAS.473.4077P}
A.~{Pillepich}, V.~{Springel}, D.~{Nelson}, S.~{Genel}, J.~{Naiman},
  R.~{Pakmor}, L.~{Hernquist}, P.~{Torrey}, M.~{Vogelsberger}, R.~{Weinberger},
  and F.~{Marinacci}, {\it {Simulating galaxy formation with the IllustrisTNG
  model}},  {\em \mnras} {\bf 473} (Jan., 2018) 4077--4106,
  [\href{http://arxiv.org/abs/1703.02970}{{\tt arXiv:1703.02970}}].

\bibitem{2018MNRAS.475..676S}
V.~{Springel}, R.~{Pakmor}, A.~{Pillepich}, R.~{Weinberger}, D.~{Nelson},
  L.~{Hernquist}, M.~{Vogelsberger}, S.~{Genel}, P.~{Torrey}, F.~{Marinacci},
  and J.~{Naiman}, {\it {First results from the IllustrisTNG simulations:
  matter and galaxy clustering}},  {\em \mnras} {\bf 475} (Mar., 2018)
  676--698, [\href{http://arxiv.org/abs/1707.03397}{{\tt arXiv:1707.03397}}].

\bibitem{2023MNRAS.524.2539P}
R.~{Pakmor}, V.~{Springel}, J.~P. {Coles}, T.~{Guillet}, C.~{Pfrommer},
  S.~{Bose}, M.~{Barrera}, A.~M. {Delgado}, F.~{Ferlito}, C.~{Frenk},
  B.~{Hadzhiyska}, C.~{Hern{\'a}ndez-Aguayo}, L.~{Hernquist}, R.~{Kannan}, and
  S.~D.~M. {White}, {\it {The MillenniumTNG Project: the hydrodynamical full
  physics simulation and a first look at its galaxy clusters}},  {\em \mnras}
  {\bf 524} (Sept., 2023) 2539--2555,
  [\href{http://arxiv.org/abs/2210.10060}{{\tt arXiv:2210.10060}}].

\bibitem{Schaye:2023jqv}
J.~Schaye et~al., {\it {The FLAMINGO project: cosmological hydrodynamical
  simulations for large-scale structure and galaxy cluster surveys}},  {\em
  Mon. Not. Roy. Astron. Soc.} {\bf 526} (2023), no.~4 4978--5020,
  [\href{http://arxiv.org/abs/2306.04024}{{\tt arXiv:2306.04024}}].

\bibitem{Kugel:2023wte}
R.~Kugel et~al., {\it {FLAMINGO: calibrating large cosmological hydrodynamical
  simulations with machine learning}},  {\em Mon. Not. Roy. Astron. Soc.} {\bf
  526} (2023), no.~4 6103--6127, [\href{http://arxiv.org/abs/2306.05492}{{\tt
  arXiv:2306.05492}}].

\bibitem{Bahe:2025kua}
Y.~M. Bahe and P.~Jablonka, {\it {Galaxies in the simulated cosmic web: I.
  Filament identification and their properties}},
  \href{http://arxiv.org/abs/2502.06484}{{\tt arXiv:2502.06484}}.

\bibitem{Galarraga-Espinosa:2021kbe}
D.~Gal\'arraga-Espinosa, M.~Langer, and N.~Aghanim, {\it {Relative distribution
  of dark matter, gas, and stars around cosmic filaments in the IllustrisTNG
  simulation}},  {\em Astron. Astrophys.} {\bf 661} (2022) A115,
  [\href{http://arxiv.org/abs/2109.06198}{{\tt arXiv:2109.06198}}].

\bibitem{Hahn:2007ui}
O.~Hahn, C.~M. Carollo, C.~Porciani, and A.~Dekel, {\it {The Evolution of Dark
  Matter Halo Properties in Clusters, Filaments, Sheets and Voids}},  {\em Mon.
  Not. Roy. Astron. Soc.} {\bf 381} (2007) 41,
  [\href{http://arxiv.org/abs/0704.2595}{{\tt arXiv:0704.2595}}].

\bibitem{Codis:2012ep}
S.~Codis, C.~Pichon, J.~Devriendt, A.~Slyz, D.~Pogosyan, Y.~Dubois, and
  T.~Sousbie, {\it {Connecting the cosmic web to the spin of dark halos:
  implications for galaxy formation}},  {\em Mon. Not. Roy. Astron. Soc.} {\bf
  427} (2012) 3320, [\href{http://arxiv.org/abs/1201.5794}{{\tt
  arXiv:1201.5794}}].

\bibitem{Trowland:2012pi}
H.~E. Trowland, G.~F. Lewis, and J.~Bland-Hawthorn, {\it {The cosmic history of
  the spin of dark matter haloes within the large scale structure}},  {\em
  Astrophys. J.} {\bf 762} (2013) 72,
  [\href{http://arxiv.org/abs/1201.6108}{{\tt arXiv:1201.6108}}].

\bibitem{Laigle:2013tsa}
C.~Laigle et~al., {\it {Swirling around filaments: are large-scale structure
  vortices spinning up dark halos?}},  {\em Mon. Not. Roy. Astron. Soc.} {\bf
  446} (2015) 2744--2759, [\href{http://arxiv.org/abs/1310.3801}{{\tt
  arXiv:1310.3801}}].

\bibitem{Borzyszkowski:2016kmi}
M.~Borzyszkowski, C.~Porciani, E.~Romano-Diaz, and E.~Garaldi, {\it {ZOMG
  \textendash{} I. How the cosmic web inhibits halo growth and generates
  assembly bias}},  {\em Mon. Not. Roy. Astron. Soc.} {\bf 469} (2017), no.~1
  594--611, [\href{http://arxiv.org/abs/1610.04231}{{\tt arXiv:1610.04231}}].

\bibitem{Gheller:2016knp}
C.~Gheller, F.~Vazza, M.~Br\"uggen, M.~Alpaslan, B.~W. Holwerda, A.~Hopkins,
  and J.~Liske, {\it {Evolution of cosmic filaments and of their galaxy
  population from MHD cosmological simulations}},  {\em Mon. Not. Roy. Astron.
  Soc.} {\bf 462} (2016), no.~1 448--463,
  [\href{http://arxiv.org/abs/1607.01406}{{\tt arXiv:1607.01406}}].

\bibitem{GaneshaiahVeena:2019arz}
P.~Ganeshaiah~Veena, M.~Cautun, E.~Tempel, R.~van~de Weygaert, and C.~S. Frenk,
  {\it {The Cosmic Ballet II: Spin alignment of galaxies and haloes with
  large-scale filaments in the EAGLE simulation}},  {\em Mon. Not. Roy. Astron.
  Soc.} {\bf 487} (2019), no.~2 1607--1625,
  [\href{http://arxiv.org/abs/1903.06716}{{\tt arXiv:1903.06716}}].

\bibitem{Banfi:2021nyj}
S.~Banfi, F.~Vazza, and C.~Gheller, {\it {On the alignment of haloes, filaments
  and magnetic fields in the simulated cosmic web}},  {\em Mon. Not. Roy.
  Astron. Soc.} {\bf 503} (2021), no.~3 4016--4031,
  [\href{http://arxiv.org/abs/2103.01943}{{\tt arXiv:2103.01943}}].

\bibitem{Hodgson_2021}
T.~Hodgson, F.~Vazza, M.~Johnston-Hollitt, and B.~McKinley, {\it Figaro
  simulation: Filaments \& galactic radio simulation},  {\em Publications of
  the Astronomical Society of Australia} {\bf 38} (2021).

\bibitem{May:2022gus}
S.~May and V.~Springel, {\it {The halo mass function and filaments in full
  cosmological simulations with fuzzy dark matter}},  {\em Mon. Not. Roy.
  Astron. Soc.} {\bf 524} (2023), no.~3 4256--4274,
  [\href{http://arxiv.org/abs/2209.14886}{{\tt arXiv:2209.14886}}].

\bibitem{Lokken:2022omq}
M.~Lokken, W.~Cui, J.~R. Bond, R.~Hlo\v{z}ek, N.~Murray, R.~Dav\'e, and A.~van
  Engelen, {\it {Boundless baryons: how diffuse gas contributes to anisotropic
  tSZ signal around simulated Three Hundred clusters}},  {\em Mon. Not. Roy.
  Astron. Soc.} {\bf 523} (2023), no.~1 1346--1363,
  [\href{http://arxiv.org/abs/2211.00242}{{\tt arXiv:2211.00242}}].

\bibitem{Jhee:2022lup}
H.~Jhee, H.~Song, R.~Smith, J.~Shin, I.~Park, and C.~Laigle, {\it {Tracking
  Halo Orbits and Their Mass Evolution around Large-scale Filaments}},  {\em
  Astrophys. J.} {\bf 940} (2022), no.~1 2,
  [\href{http://arxiv.org/abs/2201.09540}{{\tt arXiv:2201.09540}}].

\bibitem{Das:2023aje}
A.~Das, B.~Pandey, and S.~Sarkar, {\it {Galaxy Interactions in Filaments and
  Sheets: Insights from EAGLE Simulations}},  {\em Res. Astron. Astrophys.}
  {\bf 23} (2023), no.~11 115018, [\href{http://arxiv.org/abs/2303.16826}{{\tt
  arXiv:2303.16826}}].

\bibitem{Hasan:2023ujm}
F.~Hasan, J.~N. Burchett, D.~Hellinger, O.~Elek, D.~Nagai, S.~M. Faber, J.~R.
  Primack, D.~C. Koo, N.~Mandelker, and J.~Woo, {\it {Filaments of the Slime
  Mold Cosmic Web and How They Affect Galaxy Evolution}},  {\em Astrophys. J.}
  {\bf 970} (2024), no.~2 177, [\href{http://arxiv.org/abs/2311.01443}{{\tt
  arXiv:2311.01443}}].

\bibitem{Hunde:2024wic}
F.~M. Hunde, O.~Newton, W.~A. Hellwing, M.~Bilicki, and K.~Naidoo, {\it {Caught
  in the cosmic web: environmental effects on subhalo abundance and internal
  density profiles}},  \href{http://arxiv.org/abs/2409.09226}{{\tt
  arXiv:2409.09226}}.

\bibitem{Martizzi:2018iik}
D.~Martizzi et~al., {\it {Baryons in the Cosmic Web of IllustrisTNG
  \textendash{} I: gas in knots, filaments, sheets, and voids}},  {\em Mon.
  Not. Roy. Astron. Soc.} {\bf 486} (2019), no.~3 3766--3787,
  [\href{http://arxiv.org/abs/1810.01883}{{\tt arXiv:1810.01883}}].

\bibitem{Gouin:2022kcr}
C.~Gouin, S.~Gallo, and N.~Aghanim, {\it {Gas distribution from clusters to
  filaments in IllustrisTNG}},  {\em Astron. Astrophys.} {\bf 664} (2022) A198,
  [\href{http://arxiv.org/abs/2201.00593}{{\tt arXiv:2201.00593}}].

\bibitem{Li:2022zwu}
R.~Li et~al., {\it {ELUCID. VII. Using Constrained Hydro Simulations to Explore
  the Gas Component of the Cosmic Web}},  {\em Astrophys. J.} {\bf 936} (2022),
  no.~1 11, [\href{http://arxiv.org/abs/2206.08384}{{\tt arXiv:2206.08384}}].

\bibitem{Vurm:2023yze}
I.~Vurm, J.~Nevalainen, S.~E. Hong, Y.~M. Bah\'e, C.~D. Vecchia, and
  P.~Hein\"am\"aki, {\it {Cosmic gas highways in C-EAGLE simulations}},  {\em
  Astron. Astrophys.} {\bf 673} (2023) A62,
  [\href{http://arxiv.org/abs/2303.03244}{{\tt arXiv:2303.03244}}].

\bibitem{Feldbrugge:2017ivf}
J.~Feldbrugge, R.~van~de Weygaert, J.~Hidding, and J.~Feldbrugge, {\it {Caustic
  Skeleton \& Cosmic Web}},  {\em JCAP} {\bf 05} (2018) 027,
  [\href{http://arxiv.org/abs/1703.09598}{{\tt arXiv:1703.09598}}].

\bibitem{Feldbrugge:2024wsb}
J.~Feldbrugge and R.~van~de Weygaert, {\it {What makes a cosmic filament? The
  dynamical origin and identity of filaments I. fundamentals in 2D}},
  \href{http://arxiv.org/abs/2405.20475}{{\tt arXiv:2405.20475}}.

\bibitem{Feldbrugge:2022npw}
J.~Feldbrugge and R.~van~de Weygaert, {\it {Cosmic web \& caustic skeleton:
  non-linear constrained realizations \textemdash{} 2D case studies}},  {\em
  JCAP} {\bf 02} (2023) 058, [\href{http://arxiv.org/abs/2212.07840}{{\tt
  arXiv:2212.07840}}].

\bibitem{Springel:2005mi}
V.~Springel, {\it {The Cosmological simulation code GADGET-2}},  {\em Mon. Not.
  Roy. Astron. Soc.} {\bf 364} (2005) 1105--1134,
  [\href{http://arxiv.org/abs/astro-ph/0505010}{{\tt astro-ph/0505010}}].

\bibitem{Power:2002sw}
C.~Power, J.~F. Navarro, A.~Jenkins, C.~S. Frenk, S.~D.~M. White, V.~Springel,
  J.~Stadel, and T.~R. Quinn, {\it {The Inner structure of Lambda CDM halos. 1.
  A Numerical convergence study}},  {\em Mon. Not. Roy. Astron. Soc.} {\bf 338}
  (2003) 14--34, [\href{http://arxiv.org/abs/astro-ph/0201544}{{\tt
  astro-ph/0201544}}].

\bibitem{Planck:2013pxb}
{\bf Planck} Collaboration, P.~A.~R. Ade et~al., {\it {Planck 2013 results.
  XVI. Cosmological parameters}},  {\em Astron. Astrophys.} {\bf 571} (2014)
  A16, [\href{http://arxiv.org/abs/1303.5076}{{\tt arXiv:1303.5076}}].

\bibitem{Brinchmann_2004}
J.~{Brinchmann}, S.~{Charlot}, S.~D.~M. {White}, C.~{Tremonti}, G.~{Kauffmann},
  T.~{Heckman}, and J.~{Brinkmann}, {\it {The physical properties of
  star-forming galaxies in the low-redshift Universe}},  {\em \mnras} {\bf 351}
  (July, 2004) 1151--1179, [\href{http://arxiv.org/abs/astro-ph/0311060}{{\tt
  astro-ph/0311060}}].

\bibitem{Taylor_2011}
E.~N. {Taylor}, A.~M. {Hopkins}, I.~K. {Baldry}, M.~J.~I. {Brown}, S.~P.
  {Driver}, L.~S. {Kelvin}, D.~T. {Hill}, A.~S.~G. {Robotham},
  J.~{Bland-Hawthorn}, D.~H. {Jones}, R.~G. {Sharp}, D.~{Thomas}, J.~{Liske},
  J.~{Loveday}, P.~{Norberg}, J.~A. {Peacock}, S.~P. {Bamford}, S.~{Brough},
  M.~{Colless}, E.~{Cameron}, C.~J. {Conselice}, S.~M. {Croom}, C.~S. {Frenk},
  M.~{Gunawardhana}, K.~{Kuijken}, R.~C. {Nichol}, H.~R. {Parkinson},
  S.~{Phillipps}, K.~A. {Pimbblet}, C.~C. {Popescu}, M.~{Prescott}, W.~J.
  {Sutherland}, R.~J. {Tuffs}, E.~{van Kampen}, and D.~{Wijesinghe}, {\it
  {Galaxy And Mass Assembly (GAMA): stellar mass estimates}},  {\em \mnras}
  {\bf 418} (Dec., 2011) 1587--1620,
  [\href{http://arxiv.org/abs/1108.0635}{{\tt arXiv:1108.0635}}].

\bibitem{Sousbie:2010fp}
T.~Sousbie, {\it {The persistent cosmic web and its filamentary structure I:
  Theory and implementation}},  {\em Mon. Not. Roy. Astron. Soc.} {\bf 414}
  (2011) 350, [\href{http://arxiv.org/abs/1009.4015}{{\tt arXiv:1009.4015}}].

\bibitem{Sousbie:2010fn}
T.~Sousbie, C.~Pichon, and H.~Kawahara, {\it {The persistent cosmic web and its
  filamentary structure II: Illustrations}},  {\em Mon. Not. Roy. Astron. Soc.}
  {\bf 414} (2011) 384, [\href{http://arxiv.org/abs/1009.4014}{{\tt
  arXiv:1009.4014}}].

\bibitem{Galarraga-Espinosa:2023zmv}
D.~Gal\'arraga-Espinosa et~al., {\it {Evolution of cosmic filaments in the MTNG
  simulation}},  {\em Astron. Astrophys.} {\bf 684} (2024) A63,
  [\href{http://arxiv.org/abs/2309.08659}{{\tt arXiv:2309.08659}}].

\bibitem{Schaap_2000}
W.~E. {Schaap} and R.~{van de Weygaert}, {\it {Continuous fields and discrete
  samples: reconstruction through Delaunay tessellations}},  {\em \aap} {\bf
  363} (Nov., 2000) L29--L32,
  [\href{http://arxiv.org/abs/astro-ph/0011007}{{\tt astro-ph/0011007}}].

\bibitem{vandeWeygaert:2007ze}
R.~van~de Weygaert and W.~Schaap, {\it {The Cosmic Web: Geometric Analysis}},
  {\em Lect. Notes Phys.} {\bf 665} (2009) 291,
  [\href{http://arxiv.org/abs/0708.1441}{{\tt arXiv:0708.1441}}].

\bibitem{Galarraga-Espinosa:2020rhp}
D.~Gal\'arraga-Espinosa, N.~Aghanim, M.~Langer, C.~Gouin, and N.~Malavasi, {\it
  {Populations of filaments from the distribution of galaxies in numerical
  simulations}},  {\em Astron. Astrophys.} {\bf 641} (2020) A173,
  [\href{http://arxiv.org/abs/2003.09697}{{\tt arXiv:2003.09697}}].

\bibitem{Libeskind:2017tun}
N.~I. Libeskind et~al., {\it {Tracing the cosmic web}},  {\em Mon. Not. Roy.
  Astron. Soc.} {\bf 473} (2018), no.~1 1195--1217,
  [\href{http://arxiv.org/abs/1705.03021}{{\tt arXiv:1705.03021}}].

\bibitem{Cornwell:2023ntz}
D.~J. Cornwell, U.~Kuchner, M.~E. Gray, A.~Arag\'on-Salamanca, F.~R. Pearce,
  W.~Cui, and A.~Knebe, {\it {The localization of galaxy groups in close
  proximity to galaxy clusters using cosmic web nodes}},  {\em Mon. Not. Roy.
  Astron. Soc.} {\bf 527} (2023), no.~1 23--34,
  [\href{http://arxiv.org/abs/2310.11268}{{\tt arXiv:2310.11268}}].

\bibitem{hasan2024}
F.~Hasan, J.~N. Burchett, D.~Hellinger, O.~Elek, D.~Nagai, S.~M. Faber, J.~R.
  Primack, D.~C. Koo, N.~Mandelker, and J.~Woo, {\it Filaments of the slime
  mold cosmic web and how they affect galaxy evolution},  2024.

\bibitem{Wang:2024qej}
W.~Wang et~al., {\it {The boundary of cosmic filaments}},  {\em Mon. Not. Roy.
  Astron. Soc.} {\bf 532} (2024), no.~4 4604--4615,
  [\href{http://arxiv.org/abs/2402.11678}{{\tt arXiv:2402.11678}}].

\bibitem{Cavaliere:1976tx}
A.~Cavaliere and R.~Fusco-Femiano, {\it {X-rays from hot plasma in clusters of
  galaxies}},  {\em Astron. Astrophys.} {\bf 49} (1976) 137--144.

\bibitem{Ettori:2013tka}
S.~Ettori, A.~Donnarumma, E.~Pointecouteau, T.~H. Reiprich, S.~Giodini,
  L.~Lovisari, and R.~W. Schmidt, {\it {Mass profiles of Galaxy Clusters from
  X-ray analysis}},  {\em Space Sci. Rev.} {\bf 177} (2013) 119--154,
  [\href{http://arxiv.org/abs/1303.3530}{{\tt arXiv:1303.3530}}].

\bibitem{Navarro:1996gj}
J.~F. Navarro, C.~S. Frenk, and S.~D.~M. White, {\it {A Universal density
  profile from hierarchical clustering}},  {\em Astrophys. J.} {\bf 490} (1997)
  493--508, [\href{http://arxiv.org/abs/astro-ph/9611107}{{\tt
  astro-ph/9611107}}].

\bibitem{longair2011high}
M.~S. Longair, {\em High energy astrophysics}.
\newblock Cambridge university press, 2011.

\bibitem{Wayth15}
R.~B. {Wayth}, E.~{Lenc}, M.~E. {Bell}, J.~R. {Callingham}, K.~S.
  {Dwarakanath}, T.~M.~O. {Franzen}, B.-Q. {For}, B.~{Gaensler}, P.~{Hancock},
  L.~{Hindson}, N.~{Hurley-Walker}, C.~A. {Jackson}, M.~{Johnston-Hollitt},
  A.~D. {Kapi{\'n}ska}, B.~{McKinley}, J.~{Morgan}, A.~R. {Offringa},
  P.~{Procopio}, L.~{Staveley-Smith}, C.~{Wu}, Q.~{Zheng}, C.~M. {Trott},
  G.~{Bernardi}, J.~D. {Bowman}, F.~{Briggs}, R.~J. {Cappallo}, B.~E. {Corey},
  A.~A. {Deshpande}, D.~{Emrich}, R.~{Goeke}, L.~J. {Greenhill}, B.~J.
  {Hazelton}, D.~L. {Kaplan}, J.~C. {Kasper}, E.~{Kratzenberg}, C.~J.
  {Lonsdale}, M.~J. {Lynch}, S.~R. {McWhirter}, D.~A. {Mitchell}, M.~F.
  {Morales}, E.~{Morgan}, D.~{Oberoi}, S.~M. {Ord}, T.~{Prabu}, A.~E.~E.
  {Rogers}, A.~{Roshi}, N.~U. {Shankar}, K.~S. {Srivani}, R.~{Subrahmanyan},
  S.~J. {Tingay}, M.~{Waterson}, R.~L. {Webster}, A.~R. {Whitney},
  A.~{Williams}, and C.~L. {Williams}, {\it {GLEAM: The GaLactic and
  Extragalactic All-Sky MWA Survey}},  {\em Publications of the Astronomical
  Society of Australia} {\bf 32} (June, 2015) e025,
  [\href{http://arxiv.org/abs/1505.06041}{{\tt arXiv:1505.06041}}].

\bibitem{Hurley-Walker17}
N.~{Hurley-Walker}, J.~R. {Callingham}, P.~J. {Hancock}, T.~M.~O. {Franzen},
  L.~{Hindson}, A.~D. {Kapi{\'n}ska}, J.~{Morgan}, A.~R. {Offringa}, R.~B.
  {Wayth}, C.~{Wu}, Q.~{Zheng}, T.~{Murphy}, M.~E. {Bell}, K.~S. {Dwarakanath},
  B.~{For}, B.~M. {Gaensler}, M.~{Johnston-Hollitt}, E.~{Lenc}, P.~{Procopio},
  L.~{Staveley-Smith}, R.~{Ekers}, J.~D. {Bowman}, F.~{Briggs}, R.~J.
  {Cappallo}, A.~A. {Deshpande}, L.~{Greenhill}, B.~J. {Hazelton}, D.~L.
  {Kaplan}, C.~J. {Lonsdale}, S.~R. {McWhirter}, D.~A. {Mitchell}, M.~F.
  {Morales}, E.~{Morgan}, D.~{Oberoi}, S.~M. {Ord}, T.~{Prabu}, N.~U.
  {Shankar}, K.~S. {Srivani}, R.~{Subrahmanyan}, S.~J. {Tingay}, R.~L.
  {Webster}, A.~{Williams}, and C.~L. {Williams}, {\it {GaLactic and
  Extragalactic All-sky Murchison Widefield Array (GLEAM) survey - I. A
  low-frequency extragalactic catalogue}},  {\em \mnras} {\bf 464} (Jan, 2017)
  1146--1167, [\href{http://arxiv.org/abs/1610.08318}{{\tt arXiv:1610.08318}}].

\bibitem{Eastwood18}
M.~W. {Eastwood}, M.~M. {Anderson}, R.~M. {Monroe}, G.~{Hallinan}, B.~R.
  {Barsdell}, S.~A. {Bourke}, M.~A. {Clark}, S.~W. {Ellingson}, J.~{Dowell},
  H.~{Garsden}, L.~J. {Greenhill}, J.~M. {Hartman}, J.~{Kocz}, T.~J.~W.
  {Lazio}, D.~C. {Price}, F.~K. {Schinzel}, G.~B. {Taylor}, H.~K. {Vedantham},
  Y.~{Wang}, and D.~P. {Woody}, {\it {The Radio Sky at Meter Wavelengths:
  m-mode Analysis Imaging with the OVRO-LWA}},  {\em \aj} {\bf 156} (July,
  2018) 32, [\href{http://arxiv.org/abs/1711.00466}{{\tt arXiv:1711.00466}}].

\bibitem{Cowan:2010js}
G.~Cowan, K.~Cranmer, E.~Gross, and O.~Vitells, {\it {Asymptotic formulae for
  likelihood-based tests of new physics}},  {\em Eur. Phys. J. C} {\bf 71}
  (2011) 1554, [\href{http://arxiv.org/abs/1007.1727}{{\tt arXiv:1007.1727}}].
  [Erratum: Eur.Phys.J.C 73, 2501 (2013)].

\bibitem{Janish:2023kvi}
R.~Janish and E.~Pinetti, {\it {Hunting Dark Matter Lines in the Infrared
  Background with the James Webb Space Telescope}},  {\em Phys. Rev. Lett.}
  {\bf 134} (2025), no.~7 071002, [\href{http://arxiv.org/abs/2310.15395}{{\tt
  arXiv:2310.15395}}].

\bibitem{Pinetti:2025owq}
E.~Pinetti, {\it {First constraints on QCD axion dark matter using James Webb
  Space Telescope observations}},  \href{http://arxiv.org/abs/2503.11753}{{\tt
  arXiv:2503.11753}}.

\bibitem{Brown:2017dwx}
S.~Brown, T.~Vernstrom, E.~Carretti, K.~Dolag, B.~M. Gaensler,
  L.~Staveley-Smith, G.~Bernardi, M.~Haverkorn, M.~Kesteven, and S.~Poppi, {\it
  {Limiting Magnetic Fields in the Cosmic Web with Diffuse Radio Emission}},
  {\em Mon. Not. Roy. Astron. Soc.} {\bf 468} (2017), no.~4 4246--4253,
  [\href{http://arxiv.org/abs/1703.07829}{{\tt arXiv:1703.07829}}].

\bibitem{Vacca:2018rta}
V.~Vacca et~al., {\it {Observations of a nearby filament of galaxy clusters
  with the Sardinia Radio Telescope}},  {\em Mon. Not. Roy. Astron. Soc.} {\bf
  479} (2018), no.~1 776--806, [\href{http://arxiv.org/abs/1804.09199}{{\tt
  arXiv:1804.09199}}].

\bibitem{OSullivan:2018shr}
S.~P. O'Sullivan et~al., {\it {The intergalactic magnetic field probed by a
  giant radio galaxy}},  {\em Astron. Astrophys.} {\bf 622} (2019) A16,
  [\href{http://arxiv.org/abs/1811.07934}{{\tt arXiv:1811.07934}}].

\bibitem{Vernstrom:2019gjr}
T.~Vernstrom, B.~Gaensler, L.~Rudnick, and H.~Andernach, {\it {Differences in
  Faraday Rotation Between Adjacent Extragalactic Radio Sources as a Probe of
  Cosmic Magnetic Fields}},  {\em Astrophys. J.} {\bf 878} (2019), no.~2 92,
  [\href{http://arxiv.org/abs/1905.02410}{{\tt arXiv:1905.02410}}].

\bibitem{OSullivan:2020pll}
S.~P. O'Sullivan et~al., {\it {New constraints on the magnetization of the
  cosmic web using LOFAR Faraday rotation observations}},  {\em Mon. Not. Roy.
  Astron. Soc.} {\bf 495} (2020), no.~3 2607--2619,
  [\href{http://arxiv.org/abs/2002.06924}{{\tt arXiv:2002.06924}}].

\bibitem{Locatelli:2021byc}
N.~Locatelli, F.~Vazza, A.~Bonafede, S.~Banfi, G.~Bernardi, C.~Gheller,
  A.~Botteon, and T.~Shimwell, {\it {New constraints on the magnetic field in
  cosmic web filaments}},  {\em Astron. Astrophys.} {\bf 652} (2021) A80,
  [\href{http://arxiv.org/abs/2101.06051}{{\tt arXiv:2101.06051}}].

\bibitem{Amaral:2021mly}
A.~D. Amaral, T.~Vernstrom, and B.~M. Gaensler, {\it {Constraints on
  Large-Scale Magnetic Fields in the Intergalactic Medium Using
  Cross-Correlation Methods}},  {\em Mon. Not. Roy. Astron. Soc.} {\bf 503}
  (2021), no.~2 2913--2926, [\href{http://arxiv.org/abs/2102.11312}{{\tt
  arXiv:2102.11312}}].

\bibitem{Carretti:2022tbj}
E.~Carretti et~al., {\it {Magnetic field strength in cosmic web filaments}},
  {\em Mon. Not. Roy. Astron. Soc.} {\bf 512} (2022), no.~1 945--959,
  [\href{http://arxiv.org/abs/2202.04607}{{\tt arXiv:2202.04607}}].

\bibitem{Hoang:2023uyx}
D.~N. Hoang et~al., {\it {A search for intercluster filaments with LOFAR and
  eROSITA}},  {\em Mon. Not. Roy. Astron. Soc.} {\bf 523} (2023), no.~4
  6320--6335, [\href{http://arxiv.org/abs/2306.03904}{{\tt arXiv:2306.03904}}].

\bibitem{Anderson:2024giz}
C.~S. Anderson et~al., {\it {Probing the magnetized gas distribution in galaxy
  groups and the cosmic web with POSSUM Faraday rotation measures}},  {\em Mon.
  Not. Roy. Astron. Soc.} {\bf 533} (2024), no.~4 4068--4080,
  [\href{http://arxiv.org/abs/2407.20325}{{\tt arXiv:2407.20325}}].

\bibitem{Carretti:2024bcf}
E.~Carretti, F.~Vazza, S.~P. O'Sullivan, V.~Vacca, A.~Bonafede, G.~Heald,
  C.~Horellou, S.~Mtchedlidze, and T.~Vernstrom, {\it {The nature of LOFAR
  rotation measures and new constraints on magnetic fields in cosmic filaments
  and on magnetogenesis scenarios}},  {\em Astron. Astrophys.} {\bf 693} (2025)
  A208, [\href{http://arxiv.org/abs/2411.13499}{{\tt arXiv:2411.13499}}].

\bibitem{Cirelli:2020bpc}
M.~Cirelli, N.~Fornengo, B.~J. Kavanagh, and E.~Pinetti, {\it {Integral X-ray
  constraints on sub-GeV Dark Matter}},  {\em Phys. Rev. D} {\bf 103} (2021),
  no.~6 063022, [\href{http://arxiv.org/abs/2007.11493}{{\tt
  arXiv:2007.11493}}].

\bibitem{Cirelli:2023tnx}
M.~Cirelli, N.~Fornengo, J.~Koechler, E.~Pinetti, and B.~M. Roach, {\it
  {Putting all the X in one basket: Updated X-ray constraints on sub-GeV Dark
  Matter}},  {\em JCAP} {\bf 07} (2023) 026,
  [\href{http://arxiv.org/abs/2303.08854}{{\tt arXiv:2303.08854}}].

\bibitem{Essig:2013goa}
R.~Essig, E.~Kuflik, S.~D. McDermott, T.~Volansky, and K.~M. Zurek, {\it
  {Constraining Light Dark Matter with Diffuse X-Ray and Gamma-Ray
  Observations}},  {\em JHEP} {\bf 11} (2013) 193,
  [\href{http://arxiv.org/abs/1309.4091}{{\tt arXiv:1309.4091}}].

\bibitem{Massari:2015xea}
A.~Massari, E.~Izaguirre, R.~Essig, A.~Albert, E.~Bloom, and G.~A.
  G\'omez-Vargas, {\it {Strong Optimized Conservative $Fermi$-LAT Constraints
  on Dark Matter Models from the Inclusive Photon Spectrum}},  {\em Phys. Rev.
  D} {\bf 91} (2015), no.~8 083539,
  [\href{http://arxiv.org/abs/1503.07169}{{\tt arXiv:1503.07169}}].

\bibitem{Blanco:2018esa}
C.~Blanco and D.~Hooper, {\it {Constraints on Decaying Dark Matter from the
  Isotropic Gamma-Ray Background}},  {\em JCAP} {\bf 03} (2019) 019,
  [\href{http://arxiv.org/abs/1811.05988}{{\tt arXiv:1811.05988}}].

\bibitem{Fornengo:2013xda}
N.~Fornengo, L.~Maccione, and A.~Vittino, {\it {Constraints on particle dark
  matter from cosmic-ray antiprotons}},  {\em JCAP} {\bf 04} (2014) 003,
  [\href{http://arxiv.org/abs/1312.3579}{{\tt arXiv:1312.3579}}].

\bibitem{Giesen:2015ufa}
G.~Giesen, M.~Boudaud, Y.~G\'enolini, V.~Poulin, M.~Cirelli, P.~Salati, and
  P.~D. Serpico, {\it {AMS-02 antiprotons, at last! Secondary astrophysical
  component and immediate implications for Dark Matter}},  {\em JCAP} {\bf 09}
  (2015) 023, [\href{http://arxiv.org/abs/1504.04276}{{\tt arXiv:1504.04276}}].

\bibitem{Boudaud:2016mos}
M.~Boudaud, J.~Lavalle, and P.~Salati, {\it {Novel cosmic-ray electron and
  positron constraints on MeV dark matter particles}},  {\em Phys. Rev. Lett.}
  {\bf 119} (2017), no.~2 021103, [\href{http://arxiv.org/abs/1612.07698}{{\tt
  arXiv:1612.07698}}].

\bibitem{Slatyer:2016qyl}
T.~R. Slatyer and C.-L. Wu, {\it {General Constraints on Dark Matter Decay from
  the Cosmic Microwave Background}},  {\em Phys. Rev. D} {\bf 95} (2017), no.~2
  023010, [\href{http://arxiv.org/abs/1610.06933}{{\tt arXiv:1610.06933}}].

\bibitem{Capozzi:2023xie}
F.~Capozzi, R.~Z. Ferreira, L.~Lopez-Honorez, and O.~Mena, {\it {CMB and
  Lyman-\ensuremath{\alpha} constraints on dark matter decays to photons}},
  {\em JCAP} {\bf 06} (2023) 060, [\href{http://arxiv.org/abs/2303.07426}{{\tt
  arXiv:2303.07426}}].

\bibitem{Xu:2024vdn}
C.~Xu, W.~Qin, and T.~R. Slatyer, {\it {CMB limits on decaying dark matter
  beyond the ionization threshold}},  {\em Phys. Rev. D} {\bf 110} (2024),
  no.~12 123529, [\href{http://arxiv.org/abs/2408.13305}{{\tt
  arXiv:2408.13305}}].

\bibitem{Liu:2020wqz}
H.~Liu, W.~Qin, G.~W. Ridgway, and T.~R. Slatyer, {\it
  {Lyman-\ensuremath{\alpha} constraints on cosmic heating from dark matter
  annihilation and decay}},  {\em Phys. Rev. D} {\bf 104} (2021), no.~4 043514,
  [\href{http://arxiv.org/abs/2008.01084}{{\tt arXiv:2008.01084}}].

\bibitem{Wadekar:2021qae}
D.~Wadekar and Z.~Wang, {\it {Strong constraints on decay and annihilation of
  dark matter from heating of gas-rich dwarf galaxies}},  {\em Phys. Rev. D}
  {\bf 106} (2022), no.~7 075007, [\href{http://arxiv.org/abs/2111.08025}{{\tt
  arXiv:2111.08025}}].

\bibitem{Hadzhiyska:2024qsl}
B.~Hadzhiyska et~al., {\it {Evidence for large baryonic feedback at low and
  intermediate redshifts from kinematic Sunyaev-Zel'dovich observations with
  ACT and DESI photometric galaxies}},
  \href{http://arxiv.org/abs/2407.07152}{{\tt arXiv:2407.07152}}.

\end{thebibliography}\endgroup

\end{document}